\documentclass[fleqn,usenatbib]{mnras}

\usepackage{newtxtext,newtxmath}
\usepackage[T1]{fontenc}

\DeclareRobustCommand{\VAN}[3]{#2}
\let\VANthebibliography\thebibliography
\def\thebibliography{\DeclareRobustCommand{\VAN}[3]{##3}\VANthebibliography}

\usepackage{graphicx}	% Including figure files
\usepackage{amsmath}	% Advanced maths commands
\usepackage{multirow}
\newcommand{\rxj}{RX\,J0437}

\title[A shot in the dark]{A shot in the dark: searching for dark substructures in the RX\,J0437+00 galaxy cluster}

\author[Lagattuta et al.]{
David J. Lagattuta,$^{1,2,3}$\thanks{E-mail: d.lagattuta@herts.ac.uk}
Mathilde Jauzac,$^{2,3,4,5}$
Jessica E. Doppel,$^{2,3}$
Guillaume Mahler$^{6}$
and Anna Niemiec$^{7}$
\\
$^{1}$Centre for Astrophysics Research, Department of Physics, Astronomy and Mathematics, University of Hertfordshire, Hatfield AL10 9AB, UK\\
$^{2}$Centre for Extragalactic Astronomy, Durham University, South Road, Durham DH1 3LE, UK\\
$^{3}$Institute for Computational Cosmology, Durham University, South Road, Durham DH1 3LE, UK\\
$^{4}$Astrophysics Research Centre, University of KwaZulu-Natal, Westville Campus, Durban 4041, South Africa \\
$^{5}$School of Mathematics, Statistics \& Computer Science, University of KwaZulu-Natal, Westville Campus, Durban 4041, South Africa \\
$^{6}$STAR Institute, Quartier Agora - All\'ee du six Ao\^ut, 19c B-4000 Li\`ege, Belgium\\
$^{7}$Univ. Grenoble Alpes, CNRS, Grenoble INP*, LPSC-IN2P3, 38000 Grenoble, France
}

\date{Accepted XXX. Received YYY; in original form ZZZ}

\pubyear{2026}

\begin{document}
\label{firstpage}
\pagerange{\pageref{firstpage}--\pageref{lastpage}}
\maketitle

\begin{abstract}

Obtaining a census of dark matter structures at low mass ($\leq 10^9 M_\odot$) can provide strong constraints on the nature of dark matter, though identifying such structures remains difficult. In this work, we study the galaxy cluster RX\,J0437.1+0043, taking advantage of its powerful ``exotic'' Hyperbolic-Umbilic (HU) lensing configuration to search for substructure candidates. Using a combination of high resolution imaging, IFU spectroscopy, and gravitational lensing modelling, we report on a tentative detection of a dark matter subhalo ($m_{\rm halo} = 2.25 \pm 0.94 \times 10^9 M_\odot$) near the vicinity of one of the largest HU images. We stress that this result is still preliminary and that deeper data and more advanced modelling techniques are needed to ultimately confirm this detection. Nevertheless, this work outlines the first steps towards understanding subhalo properties in dense cluster environments, developing HU cluster lenses as a potential new tool for investigating dark matter.
\end{abstract}

\begin{keywords}
galaxies: clusters: individual: RX J0437.1+0043 --  dark matter -- gravitational lensing: strong -- techniques: imaging spectroscopy
\end{keywords}

%%%%%%%%%%%%%%%%%%%%%%%%%%%%%%%%%%%%%%%%%%%%%%%%%%

%%%%%%%%%%%%%%%%% BODY OF PAPER %%%%%%%%%%%%%%%%%%

\section{Introduction}

Dark matter (DM) has been a major topic of modern astronomical research for several decades. Although it accounts for nearly 90\% of the Universe's mass budget, little is known about its physical properties; this is largely due to the fact that, to date, direct detection of a DM particle has never been achieved \citep[e.g.,][]{strigari2013,misiaszek2024}. Thus, we currently lack observational evidence of (proposed) DM characteristics such as its particle mass, its interaction cross-section, and its decay rate. Since insights into the exact nature of DM have important implications in both astrophysics and cosmology, it is imperative that we continue to investigate this elusive substance.

By interacting through gravity, DM is known to cluster and grow over cosmological time. As a result, we can indirectly detect evidence of its presence through other ways, particularly by measuring the \emph{total} mass of collapsed (or collapsing) structures. Several methods exist for ``weighing'' a DM halo, including measuring galaxy rotation curves in \ion{H}{I} gas \citep[e.g.,][]{banerjee2011, diTeodoro2023}, probing Doppler shifts in Cosmic Microwave Background photons via the Sunyaev-Zel'dovich (SZ) Effect \citep[e.g.][]{marrone2012, ade2013, zubeldia2019, huchet2024}, or tracing the temperature profile of X-ray gas in galaxy clusters \citep[e.g.,][]{sereno2017, tchernin2018, allingham2023, beauchesne2025}. However, of all the possible techniques, one method stands out as an especially efficient mass probe:~gravitational lensing. Gravitational lensing directly provides a total (dark+baryon) mass estimate for an object, without needing to invoke simplifying assumptions about the object's morphology, kinematic properties, or dynamical state. Therefore, when used in conjunction with other observables that intrinsically trace baryons (such as galaxy brightness or emission line strength), a lensing-mass estimate can cleanly disentangle the mass contributions of luminous matter from DM \citep[e.g.,][]{limousin2008, richard2010, jauzac2012, grillo2015, lagattuta2017, mahler2018, cerny2018, diego2020, niemiec2023, perera2024, rihtarsic2025}. 

At the same time, lensing is equally suited to measuring mass at both large and small scales. This allows for a simultaneous estimate of the larger-scale ``smooth'' mass component of the underlying halo, along with the distribution of smaller ``clumpy'' subhalos that exist within the primary mass. The subhalo mass function is a fundamental parameter of the cosmological model and, as such, can provide additional clues to the nature of DM. According to cosmological simulations, the standard concordance $\Lambda$-CDM model predicts that of thousands of low-mass ($< 10^9~ M_{\odot}$) DM halos should survive to the present day, as a consequence of the hierarchical build-up of the Universe's large scale structure \citep[e.g.,][]{white1991}. Conversely, alternative models such as a relativistic ``warm'' DM particle (WDM) or a nominally higher DM-cross section (self-interacting DM; SIDM) predict the existence of far fewer subhalos \citep[e.g.][]{lovell2012, zavala2019, nadler2025}. With an accurate census of low-mass substructures then, we can begin to discriminate between competing cosmological models. As many of these objects are expected to be extremely faint or even entirely dark \citep{benitez-llambay2020}, lensing once again serves as a powerful tool for their detection.  

Lensing-based efforts to detect DM subhalos and estimate their mass function have become more common in recent years, both observationally \citep[e.g.][]{vegetti2012, hezaveh2016, nightingale2024} and computationally \citep[e.g.,][]{li2017, despali2017, amorisco2022}. In all cases, the technique is largely the same: (i) identify the multiple images of a lensed background galaxy; (ii) look for distortions in either the position or brightness of individual images relative to its counterparts (as `copies' of the same source, multiple images should inherently appear identical, after accounting for lensing magnification); and finally, (iii) attempt to correct these deformations by adding a perturbing mass (i.e., a subhalo) to the lens model, in the vicinity of the distortion. However, at present, nearly all efforts in this domain have focused on galaxy-scale mass distributions, i.e., galaxy-galaxy lenses (GGLs). 

In this work, we focus on gravitational lensing in galaxy clusters, as clusters have two advantages over their galaxy-scale counterparts. First, they are physically larger and more massive than individual galaxies, which (again invoking the hierarchical nature of the Universe) implies that they contain more subhalos than GGLs \citep{doppel2025} and also span a wider mass range. This suggests that there is a higher probability of detecting subhalos in clusters, and that the statistics of these detections will be more diverse. Second, background objects being lensed by clusters lie, on average, further away from bright galaxies compared to GGLs. This means that they are less contaminated by cluster galaxy light, making it easier to identify the subtle deformations possibly highlighting the presence of nearby small-scale subhalos.

\begin{figure}
    \centering
    \includegraphics[width=\columnwidth]{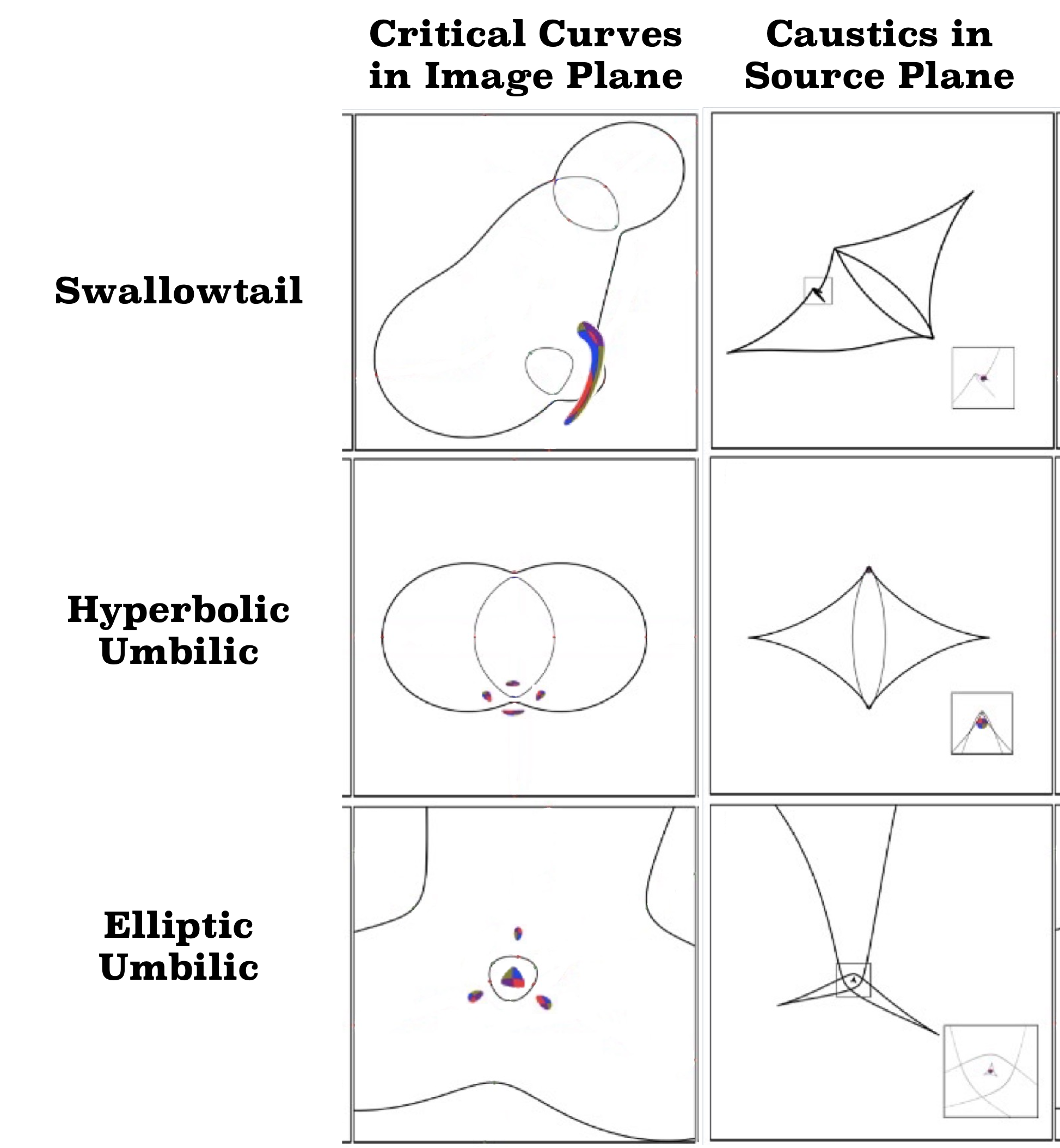}
    \caption{Examples of rare, ``exotic'' lens configurations. The left column displays the critical curves in the (observed) image plane, with the characteristic image formations and image parity shown by the multicoloured regions. The right column shows the corresponding caustics in the (demagnified, intrinsic) source plane. [Adapted from Figs.~1-3 in \citet{meena2020}.]  Lenses in exotic configurations are more sensitive to the effects of nearby substructures than more common ``stable'' lensed images, making them ideal probes to look for low-mass subhalos. The Hyperbolic-Umbilic (HU) lensing configuration seen in the middle panels are the subject of this work.} 
    \label{fig:exotics}
\end{figure}

Objects lensed by clusters can appear in a variety of configurations, ranging from common ``stable'' orientations (such as double images, quadruples, and fully connected Einstein-rings) to more ``exotic'' forms, such as swallowtail and umbilic shapes \citep[e.g.,][]{petters2021, orban2009, meena2020}; see Fig \ref{fig:exotics}. Previous work has shown that these exotic systems are more sensitive to the presence of subhalos compared to stable patterns \citep{meena2023}, with one specific exotic configuration, the hyperbolic-umbilic (HU), providing the most optimal ``net'' for catching substructures, in terms of its physical size and number of lensed images.

\begin{figure*}
    \centering
    \includegraphics[width=\linewidth]{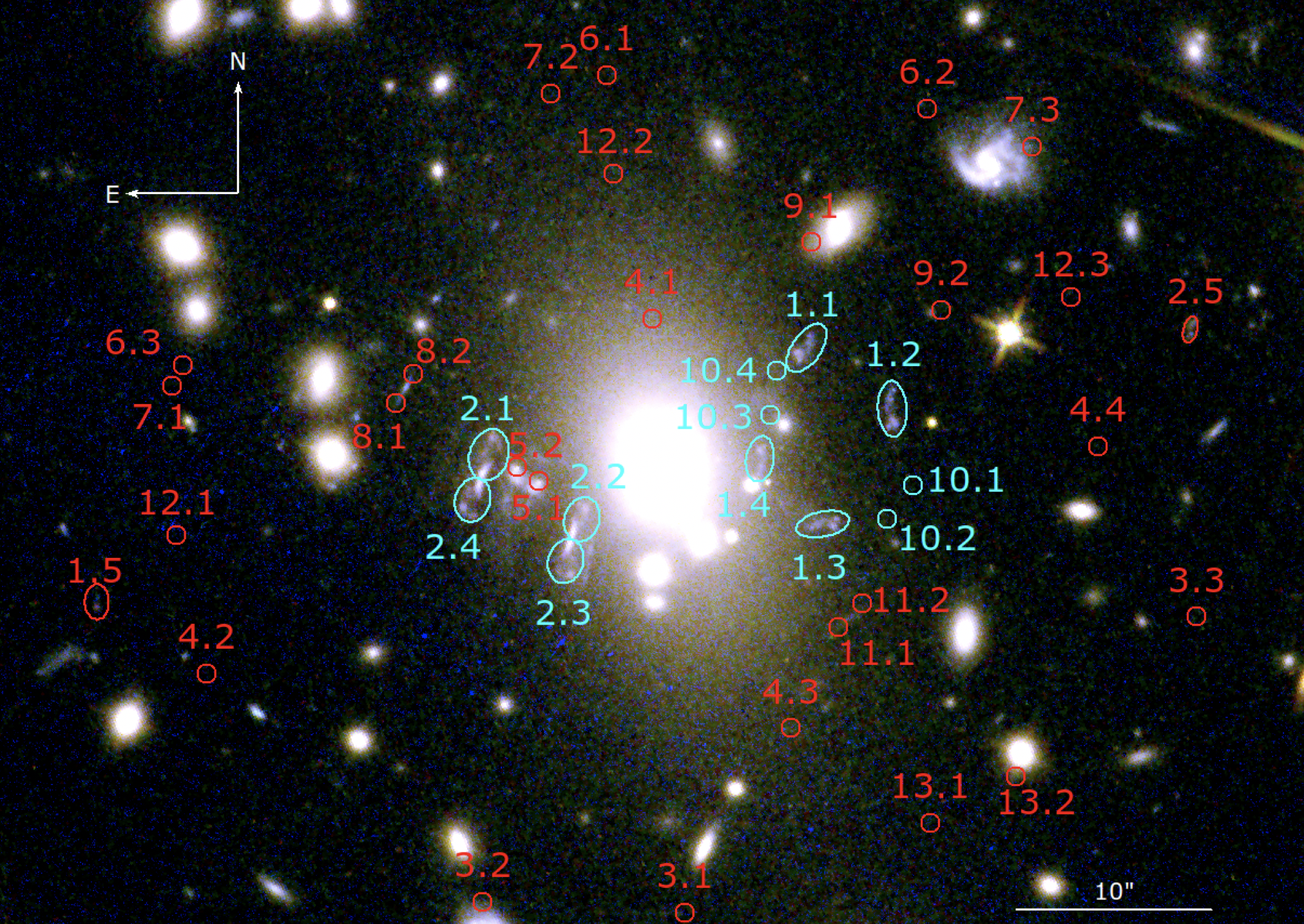}
    \caption{Spectroscopically-confirmed multiply-imaged sources in the \rxj\ field, overlaid on multi-band \emph{HST} imaging (F140W/F110W/F814W). Sources identified as Hyperbolic-Umbilic (HU) systems appear in cyan, while all other systems are shown in red. Numbering convention of all systems follows that presented in \citet{lagattuta2023}.}
    \label{fig:rxj0437}
\end{figure*}

Therefore, we here investigate the possible presence of substructures in the galaxy cluster RX\,J0437.1+0043 (\rxj; $z=0.285$). First studied in \citet{lagattuta2023}, the cluster has a largely elliptical shape with 12 spectroscopically confirmed multiple-image systems. Remarkably, despite being historically rare, \emph{three} of the \rxj~ systems appear in HU arrangements (System\,1, $z_s$ = 1.97; System\,2, $z_s$ = 2.97; and System\,10, $z_s$ = 6.02), each lying within the cluster core ($R < 100$\,kpc from the centre; see Fig.~\ref{fig:rxj0437}). High-resolution imaging from the \emph{Hubble Space Telescope} (\emph{HST}) shows that Systems\,1 and 2 are both clumpy, highly structured galaxies, while integral-field spectroscopy shows that all three systems have bright, extended emission-line features, making them ideal test cases to search for low-mass perturbers. 

This paper is organized as follows. We describe the data used in this analysis in Section\,\ref{sec:data}, then lay out the techniques used to classify features in lensed galaxies and investigate differences between their multiple images in Section\,\ref{sec:substructure}. Based on the results of this census, we create a series of strong-lensing mass models that we present in Section\,\ref{sec:modeling}, allowing us to detect perturbations in the vicinity of lensed galaxies. We discuss modelling results in Section\,\ref{sec:discussion}, presenting evidence for a possible low-mass DM substructure close to System\,1. Finally, we conclude and discuss future work in Section\,\ref{sec:conclusions}.

Throughout this work, we assume a flat cosmological model with parameters set to $\Omega_\Lambda$ = 0.7, $\Omega_{\rm M}$ = 0.3, and $H_0$ = 70\,km\,s$^{-1}$ Mpc$^{-1}$. Under these conditions, 1\,\arcsec\ covers 4.297\,kpc at the cluster redshift (z$=$0.285). Finally, unless otherwise specified, all magnitudes are presented in the AB system \citep{oke74}.

\begin{figure*}
    \centering
    \includegraphics[width=0.49\linewidth]{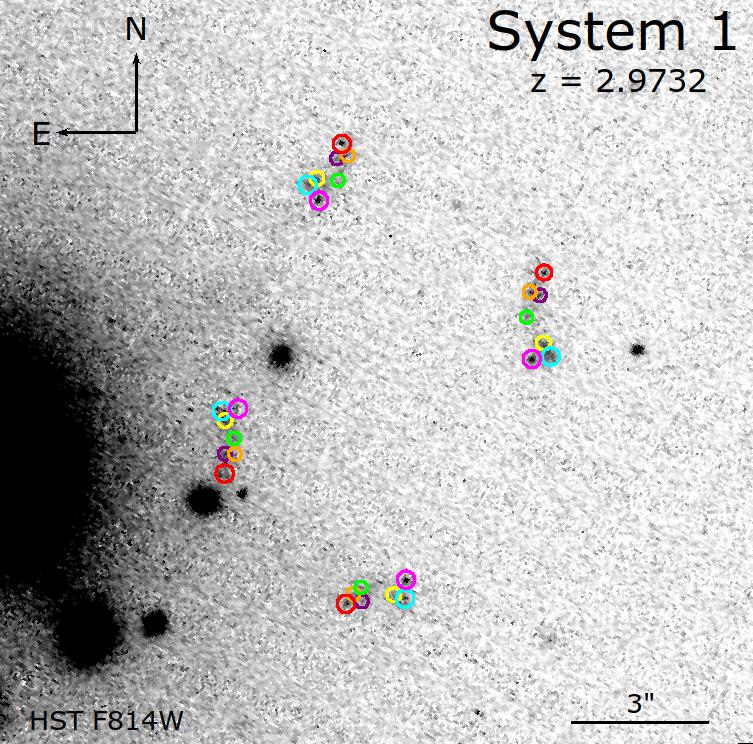}
    \includegraphics[width=0.49\linewidth]{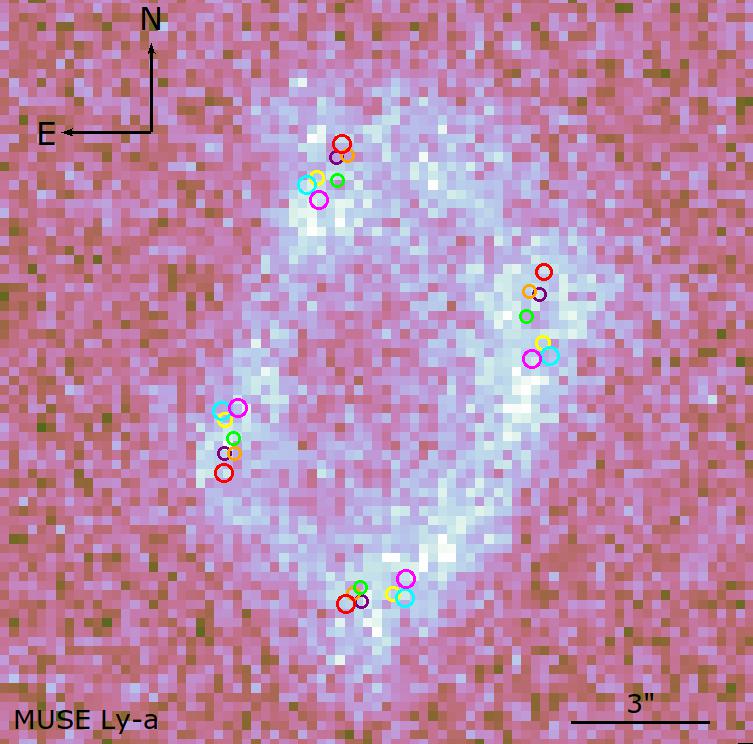}
    \includegraphics[width=0.32\linewidth]{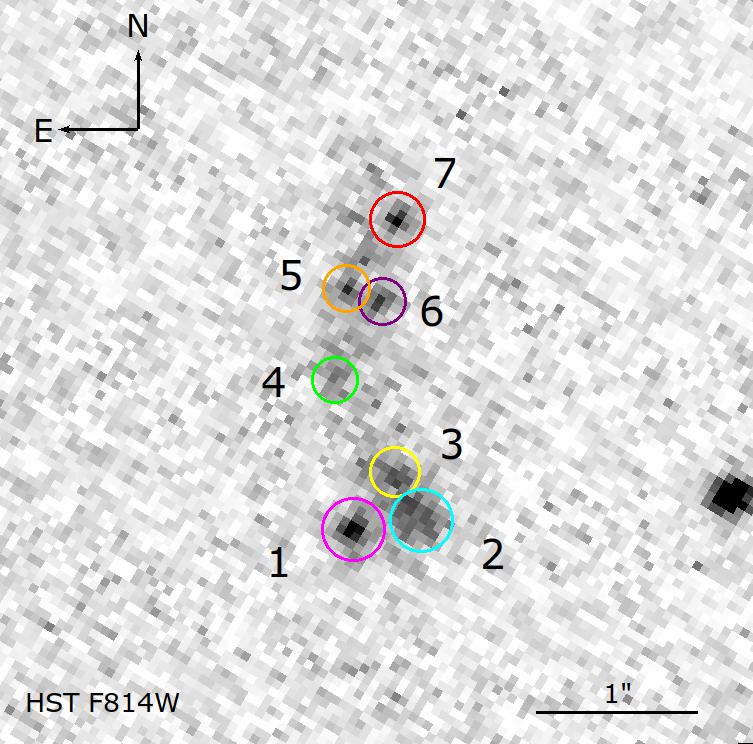}
    \includegraphics[width=0.32\linewidth]{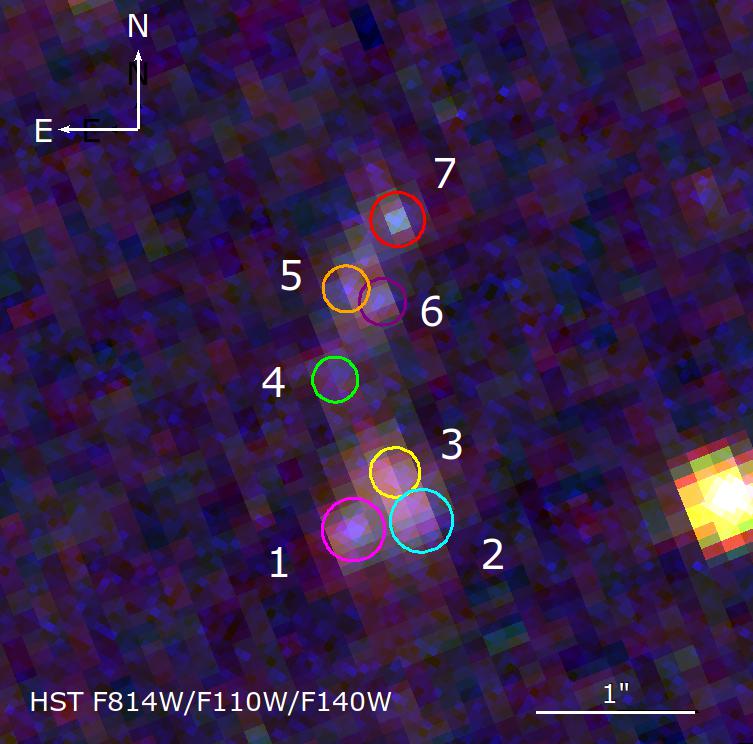}
    \includegraphics[width=0.32\linewidth]{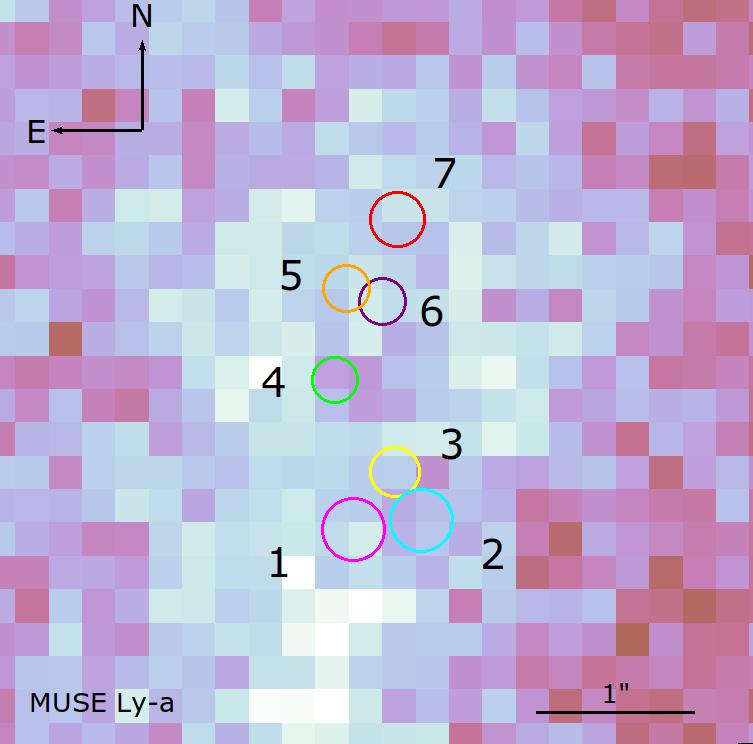}
    \caption{Multiple images of HU System\,1 ($z = 2.97$), seen in several data sets. In the top left panel, we see a greyscale image of the complete system, observed in \emph{HST}/F814W imaging, the highest resolution image available. A zoom-in of one image (Image\,1.2) is seen in the bottom left cutout. From this data we have identified seven distinct morphological features (``clumps'', labelled 1-7), which we identify with coloured apertures. A 3-band colour image of the same system is shown in the bottom middle panel. This adds colour information, but individual clumps appear less distinct, due to the lower resolution of the \emph{HST}/F110W and F140W imaging. The top right panel shows how the system appears in VLT/MUSE imaging, in this case, a complete ring structure composed of Lyman-$\alpha$ emission. A zoom in (of the same individual image) is shown in the bottom right panel. Contrasting with the \emph{HST} imaging, the MUSE view of the system shows a broad, smooth feature, rather than distinct clumps.}
    \label{fig:sys1}
\end{figure*}

\begin{figure*}
    \centering
    \includegraphics[width=0.49\linewidth]{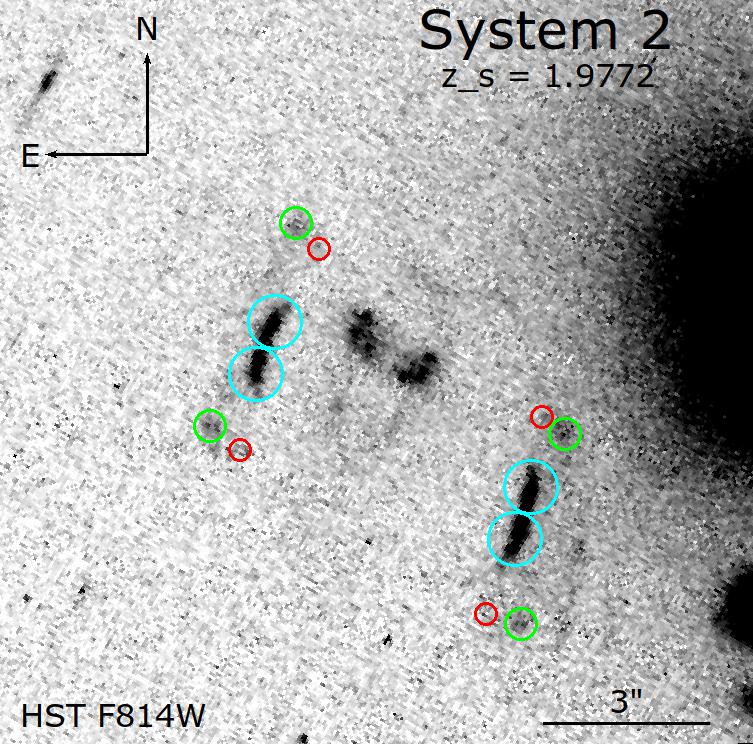}
    \includegraphics[width=0.49\linewidth]{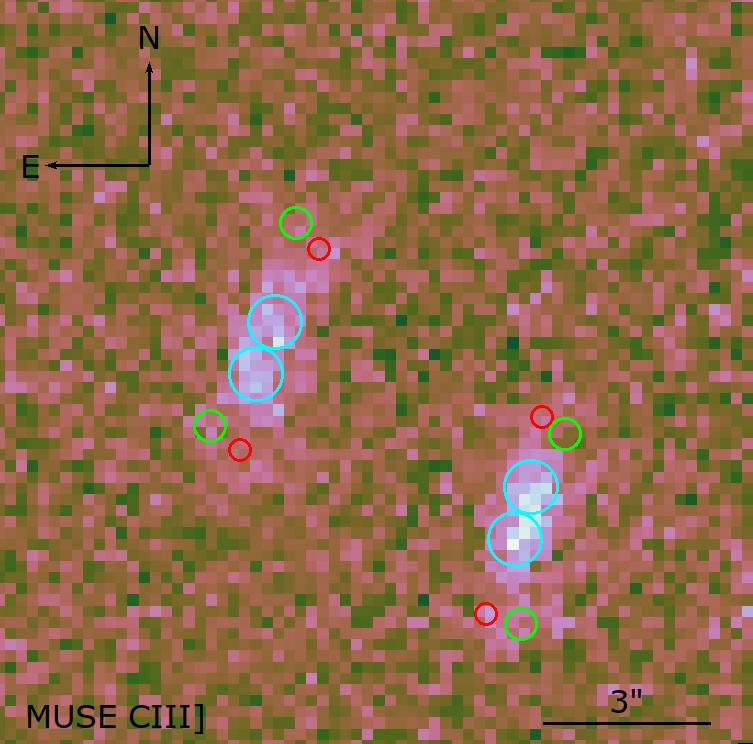}
    \includegraphics[width=0.32\linewidth]{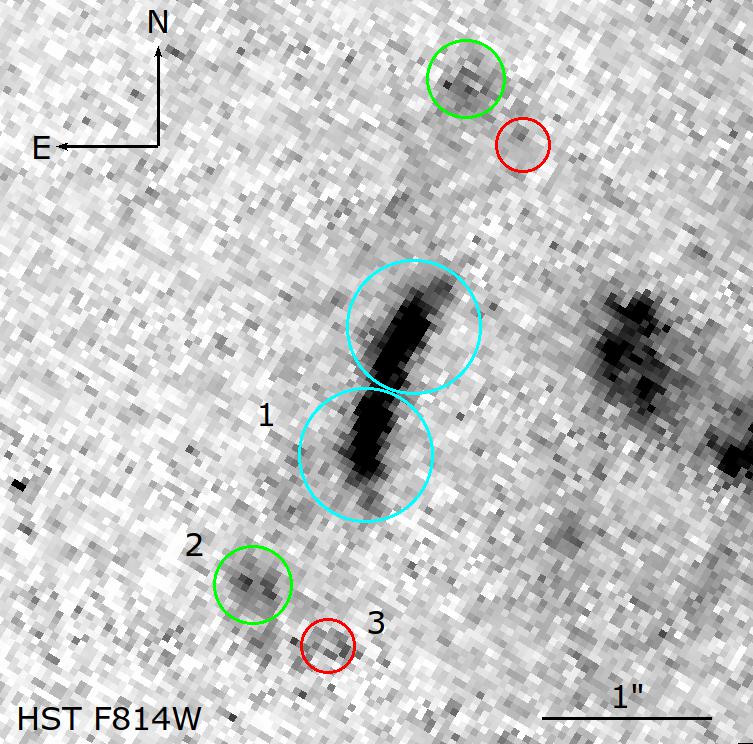}
    \includegraphics[width=0.32\linewidth]{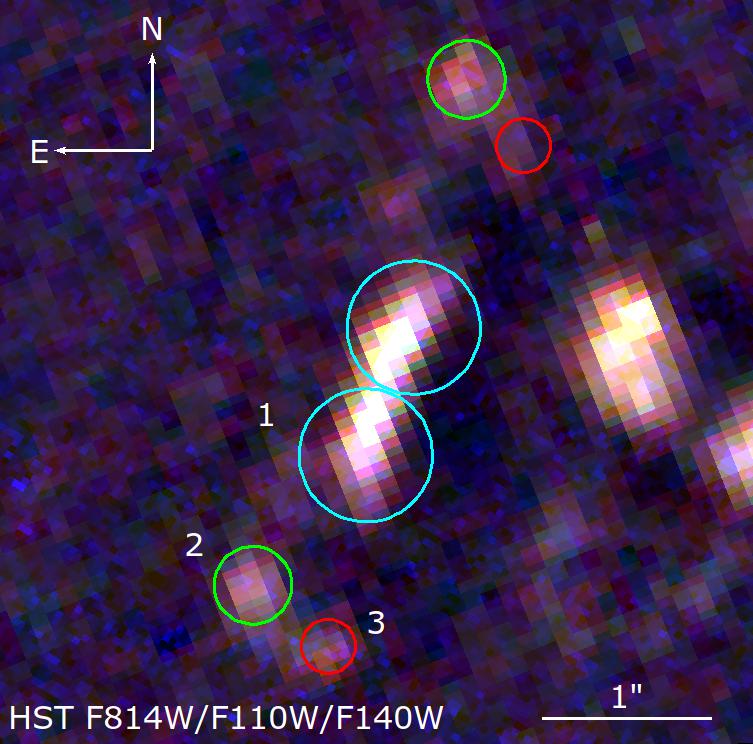}
    \includegraphics[width=0.32\linewidth]{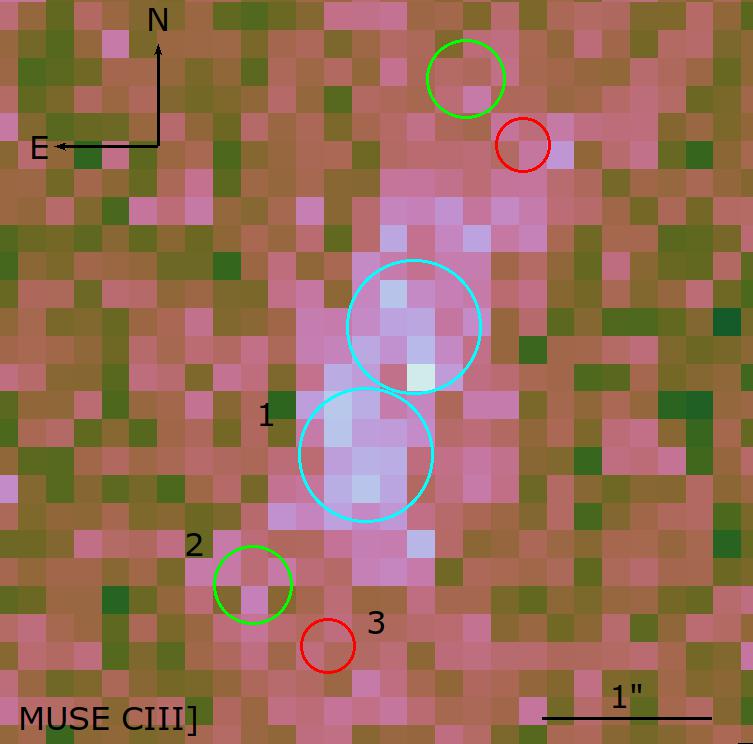}
    \caption{Multiple images of HU System\,2 ($z = 1.98$), seen in several data sets. The layout and composition of these images follows that of Fig.~\ref{fig:sys1}, though the zoomed-in image this time is the combined merging images of Images\,2.1 and 2.4. In System\,2, we identify three distinct clumps in the high-resolution imaging, which, due to their larger sizes are still distinct in the colour panel. The bright central component is readily identified in MUSE data (here CIII emission), while the two edge components are less bright.}
    \label{fig:sys2}
\end{figure*}

\begin{figure*}
    \centering
    \includegraphics[width=0.49\linewidth]{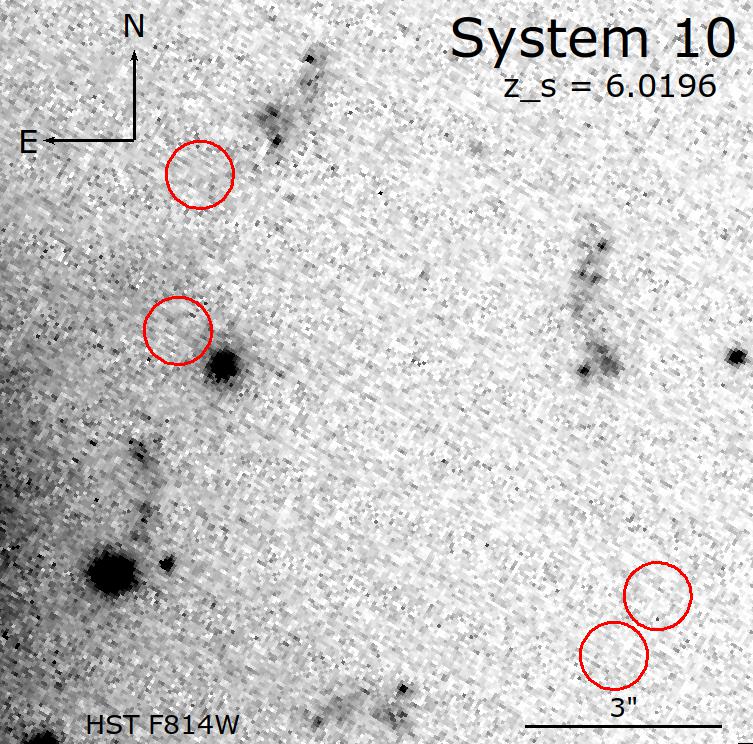}
    \includegraphics[width=0.49\linewidth]{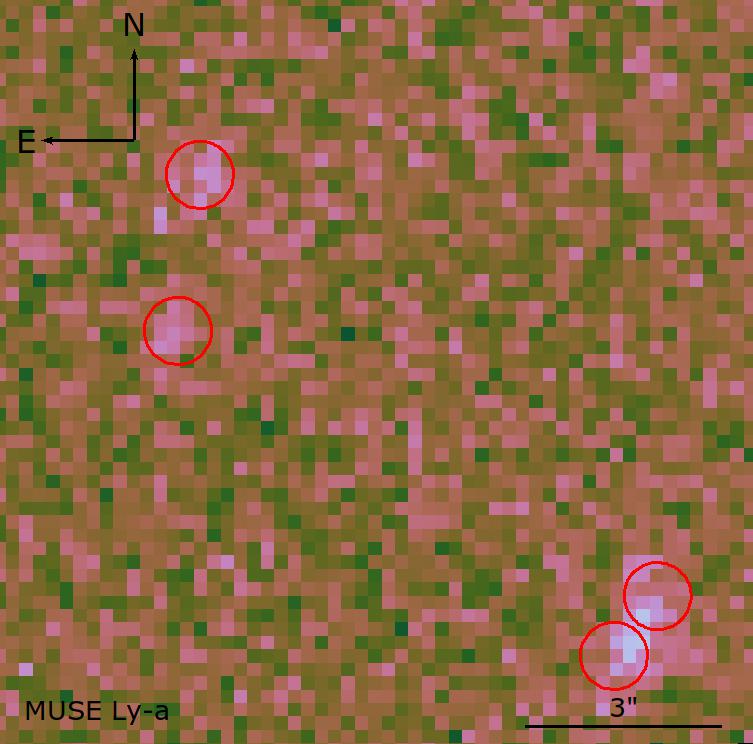}
    \includegraphics[width=0.32\linewidth]{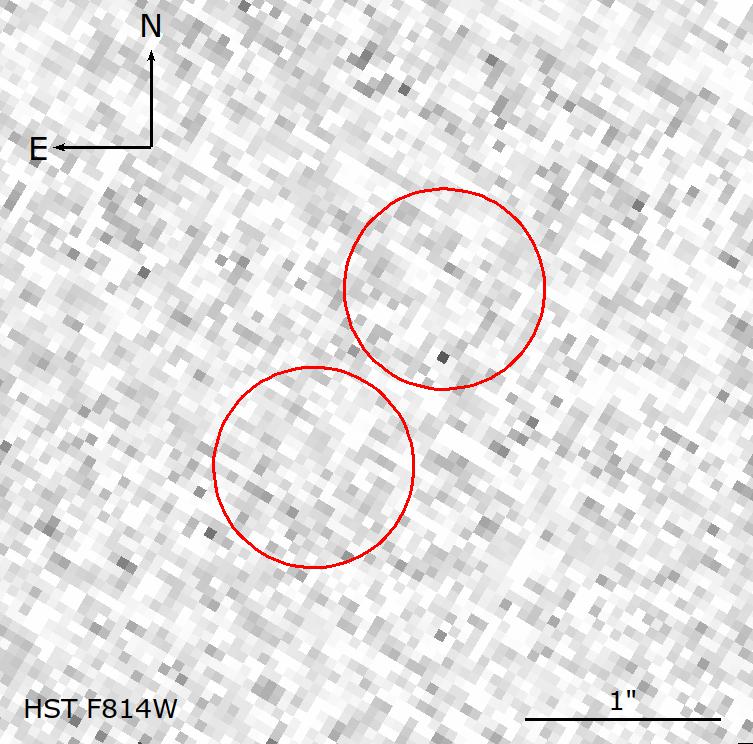}
    \includegraphics[width=0.32\linewidth]{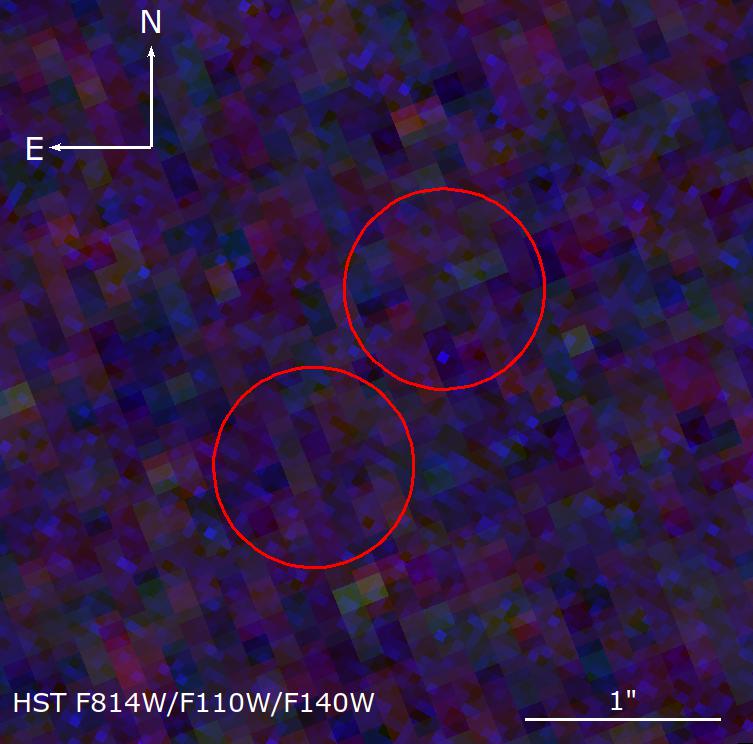}
    \includegraphics[width=0.32\linewidth]{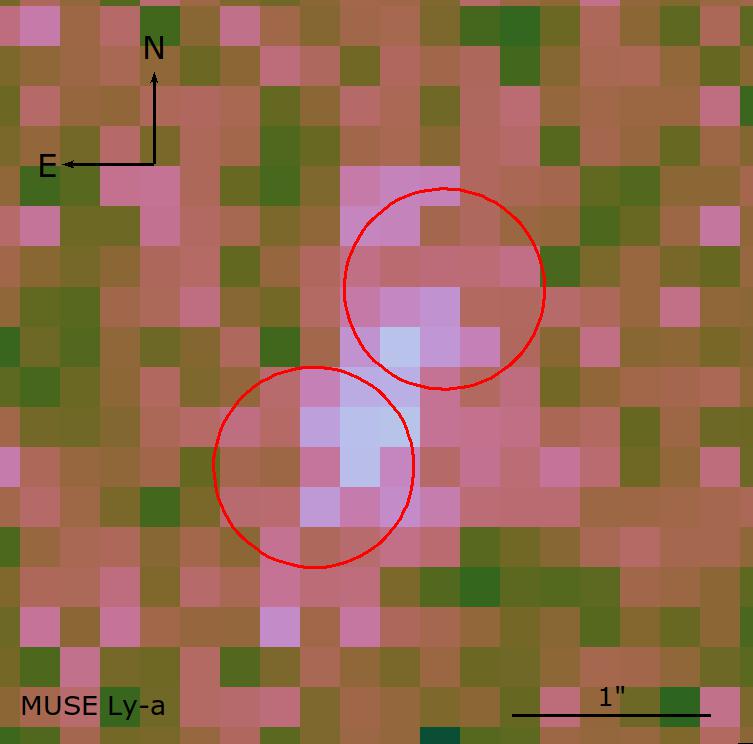}
    \caption{Multiple images of HU System\,10 ($z = 6.02$), seen in several data sets. The layout and composition of these images follows that of Fig.~\ref{fig:sys1}, though the zoomed-in image this time is the combined merging images of Images\,10.1 and 10.2. Unlike Systems\,1 and 2, System\,10 is not seen in any broadband images, having been solely identified in deep MUSE observations as a faint Lyman-$\alpha$ emitter. Future, deeper imaging may yet reveal continuum emission and more complex morphological features for this system.}
    \label{fig:sys10}
\end{figure*}

\section{Observations}
\label{sec:data}
Our analysis is based on high-resolution space-based imaging with the \emph{HST} and ground-based spectroscopy with the MUSE spectrograph at the Very Large Telescope (VLT). The reduced data used in this work are nearly identical to those presented in \citet{lagattuta2023}, as are the techniques we employ to reduce them. Full details of the data and reduction methods can be found in \citet{lagattuta2023}, but we provide a summary of the information here and highlight the minor differences in the final reduction to aid the reader.

\begin{table*}
    \centering
    \caption{Summary of the \emph{HST} observations for \rxj. Column 1 gives the instrument and the band. Columns 2 and 3 list the ID of the proposal and respective PI. Columns 4 and 5 give the total exposure time in seconds and the observation date.}
    \label{tab:hstObs}
    \begin{tabular}{c|c|c|c|c}
    \hline \hline
       Band  & PID & P.I. & Exposure time & Observation Date\\
        & & & [s] & \\
       \hline  
         ACS/F814W & 16670 & H. Ebeling & 1200 & 2021-11-21 \\
         WFC3/F110W & `` & `` & 706 & 2022-07-25 \\
         WFC3/F140W & `` & `` & 1059 & 2022-07-25 \\
         \hline
         \hline
    \end{tabular}
\end{table*}

\subsection{\emph{Hubble Space Telescope} imaging}
\emph{HST} imaging of \rxj~ exists in three broadband filters: F814W, F110W, and F140W. The F814W data were acquired on 21 November 2021 using the \emph{Advanced Camera for Surveys} (\emph{ACS}), while the F110W and F140W data were acquired on 25 July 2022 with the \emph{Wide-Field Camera 3} (\emph{WFC3}). All data were taken as part of a SNAPshot imaging campaign (GO-16670; PI: H.~Ebeling) targeting massive galaxy clusters. A summary of the relevant observational parameters is shown in Table \ref{tab:hstObs}. 

For this analysis, we use high-level data products in each band from the Mikulski Archive for Space Telescopes (MAST\footnote{\url{https://mast.stsci.edu/search/ui/}}), which provides reduced and stacked images for multi-band analysis. Reduction was performed using the publicly-available \textsc{DrizzlePAC} software\footnote{\url{https://www.stsci.edu/scientific-community/software/drizzlepac}}, which aligned individual exposures to a common World Coordinate System, eliminated systematic artefacts such as cosmic-ray strikes and hot pixels, corrected for the effects of geometric distortion, and (in the case of the \emph{ACS} image) removed photometric smearing caused by charge-transfer inefficiency tails. Unlike in \citet{lagattuta2023}, however, we applied one final correction to the stacked images: we resampled each image from its native (non-rotated) pixel scale to a common registered pixel grid (0.05\arcsec/pixel, oriented North up, East left) using SWarp\footnote{\url{https://www.astromatic.net/software/swarp/}} \citep{bertin2010}. This provided a more uniform pixel coverage between bands, making it easier to compare photometric estimates from band to band (Section\,\ref{subsec:clumpProperties}).

\begin{table*}
    \centering
    \caption{Summary of VLT/MUSE observations for \rxj. Columns 1 and 2 indicate the ID of the ESO programme and respective PI. For each pointing, we give the observation date in column 3, the istrument setup in Column 4, the total exposure time in column 5, the right ascension (R. A.) and declination (Decl.) of the center of the field of view in columns 6 and 7, and the mean FWHM of the seeing during the observations in column 7.}
    \label{tab:museObs}
    \begin{tabular}{c|c|c|c|c|c|c|c}
    \hline \hline
       ESO Programme  & P.I. & Observation Date & Instrument Setup & Exposure time & R.A. & Decl. & Seeing\\
        & & & & [s] & [J2000] & [J2000] & \arcsec\\
       \hline  
       0104.A-0801 & A.~Edge & 2020-02-15 & WFM-NOAO-N & 2910 & 69.289583 & 0.731167 & 0.49\\
       106.21AD & D.~Lagattuta & 2021-01-15 & WFM-AO-N & 2544 & 69.293333 & 0.735667 & 1.05\\
       `` & `` & `` & `` & `` & 69.286958 & 0.735611 & 1.49\\
       `` & `` & 2021-01-17 & `` & `` & 69.286917 & 0.735528 & 0.41\\
       `` & `` & 2021-01-19 & `` & `` & 69.293208 & 0.726389 & 0.70\\
       `` & `` & `` & `` & `` & 69.286792 & 0.726472 & 0.79\\
       `` & `` & `` & `` & `` & 69.293250 & 0.735750 & 0.70\\
       `` & `` & 2021-02-13 & `` & `` & 69.286917 & 0.735694 & 0.65\\
       `` & `` & 2021-02-17 & `` & `` & 69.293292 & 0.726306 & 0.83\\
         \hline
         \hline
    \end{tabular}
\end{table*}

\subsection{VLT/MUSE spectroscopy}
The VLT/MUSE data used in this work is gathered from two observing programs targeting \rxj. The first one is a single 1\,hour exposure taken on 15 February 2020 as part of the KALEIDOSCOPE Cluster survey (PID 0104.A-0801; PI A.~Edge). These data were obtained in WFM-NOAO-N mode, centred on the brightest cluster galaxy (BCG) located at ($\alpha$ = 69.289677, $\delta$ = 0.73114470). We combine this frame with a larger 1.5\arcmin$\times$ 1.5\arcmin\ mosaic of eight stacked MUSE pointings observed between 15 January and 13 February 2021 (PID 106.21AD; PI D.~Lagattuta). The mosaic is again centred on the BCG, but the data are instead acquired using Adaptive Optics (AO) corrections in WFM-AO-N mode. 

All nine data frames are reduced using the method outlined in Section\,2.2 of \citet{lagattuta2022}. This procedure largely follows the standard MUSE data reduction pipeline \citep{weilbacher20}, but includes an additional auto-calibration step to reduce flux variation between integral field unit (IFU) image slices, and applies the Zurich Astrophysical Purge (\textsc{zap}) software \citep{soto2016} to better remove sky line residuals. The combined data set spans a wavelength range from 4750\AA\ to 9350\AA, though the mosaic frames have a coverage gap between 5804\AA\ and 5967\AA\ due to the AO laser-blocking notch filter. The mean spectral resolution of the final combined data cube is $R = 3000$, and the exposure time ranges from 20352\,s in the cluster centre to 2544\,s in the outskirts.
A summary of the different MUSE observations parameters is given in Table \ref{tab:museObs}. 

\section{Substructure Analysis}
\label{sec:substructure}

The DM substructures we are interested in are considerably lower-mass and more compact compared to the larger, cluster-scale halos in which they are embedded. As a result, any visible deflections they cause will be small, extending only to $\sim$1\arcsec\, from their position \citep[e.g.,][]{he2022}. 
In practice, this means that while these substructures will not significantly alter the global lens model, their presence can create localized perturbations, subtly altering the appearance of nearby objects in the  (observed) lens plane. Given a suitable benchmark, these perturbations can be detected, particularly when looking at the (ostensibly identical) multiple images of a lensed background galaxy. Although any multiply-imaged galaxy in a cluster can, in principle, be used for such a comparison, the unique nature of HU systems makes them ideally suited for this task. 

Specifically, HU images experience a high level of magnification (up to $\sim100\times$) that is nearly isotropic, providing a larger, more uniform view of the source galaxy compared to, e.g., lensed giant arcs, which only appear strongly stretched in one direction. This is because the HU images lie close to both radial and tangential critical curves (lines of formally infinite magnification for point sources, see Fig.\,\ref{fig:exotics}). As a result, we are better able to identify unique and distinct features in the galaxy's morphology, making it easier to look for unexpected discrepancies between images. In this section, we describe the techniques used to study the properties of the HU images in \rxj~ and identify any perceptible differences between them. 

\subsection{HU galaxy images subcomponents identification}

We begin by visually inspecting each image in a given lens system, looking for distinct morphological features that can be compared across all images. We note here that, while each HU lens has five unique multiple images, our analysis focuses only on the four central images of a given system (see Fig.~\ref{fig:rxj0437}), as these are the most highly-magnified, allowing us to identify unique structural features. To make full use of all available information, we measure the object's appearance using both \emph{HST} (broadband imaging) and MUSE (narrow-band imaging, centred around known emission-line features) data. This is because each of the \rxj~ HU galaxies are young, highly star-forming systems, still actively assembling. As a result, the stellar and gas disks of the galaxy are not necessarily aligned, so the two components can provide independent sight-lines through the foreground cluster environment. Of the three \emph{HST} bands at our disposal (ACS/F814W, WFC3/F110W and WFC3/F140W), the F814W imaging has the highest resolution and the smallest PSF, providing the clearest picture for detecting distinct, clumpy subcomponents.  We therefore use the F814W data specifically for our initial feature detection, but supplement it with colour information from the other two bands. 

We identify seven distinct features in the \emph{HST} imaging of System\,1, which appear to be individual pockets of active star formation. These knots are also seen in the MUSE data -- specifically in the wavelength range covering \ion{C}{III}] emission and UV stellar continuum -- but they appear to fuse into two broader clumps (Fig.\,\ref{fig:sys1}) due to the lower resolution of MUSE. Similarly, we detect three features in System\,2, including a bright central component representing the main galaxy core, along with two additional stellar knots in the outskirts of the system. The MUSE view is again at lower resolution and, in this case, we can only confidently identify the bright central core, though the separation between individual multiply-imaged components is still distinct (Fig.\,\ref{fig:sys2}). Finally, while we attempt to identify features of System\,10, we see no evidence of galaxy continuum in any \emph{HST} bands, though given the shallowness of the exposures (and a redshift $z \sim 6$) this is perhaps unsurprising. Instead, System\,10 appears only as a single, nearly point-like feature in the MUSE data, with the two Northern-most images appearing as distinct objects, and the two Southern components appearing as a merging pair. Unlike the other HU galaxies, the images in this system appear only as a Lyman-$\alpha$ (Ly$\alpha$) emission line, with no discernible continuum emission elsewhere in the data cube (Fig.\,\ref{fig:sys10}). Given the lower resolution of the MUSE-only detection, coupled with the lack of colour information and difficulty in identifying unique features in each component -- the smoothness of the Ly$\alpha$ flux in the merging pair prevents a distinct separation of images -- we decided to focus solely on Systems\,1 and 2 in the remaining of this analysis. However, we note that it would be interesting to revisit System\,10 in the future using deeper \emph{HST} or \emph{JWST} observations.

\subsection{HU galaxy images subcomponents properties}
\label{subsec:clumpProperties}

After identifying individual clumps, we characterize some of their key physical properties, making separate measurements in each HU image, to look for cross-image discrepancies. We specifically focus on two features for each clump: its centroid position and total (multi-band) flux, as differences between images in these quantities (known as astrometric and flux-ratio anomalies, respectively) can indicate the presence of a nearby low-mass perturber.

We first attempt to make these measurements using \texttt{SourceExtractor} \citep{sextractor1996}, in order to automatically detect each clump's centroid and standardize the aperture used to measure its flux. To maintain consistency between measurements in different bands, we run \texttt{SourceExtractor} in dual image mode, using the F814W band as the detection image, and a second filter (including F814W itself) as the photometric image. However, after several attempts (including modifications to our choice of \texttt{Source Extractor} detection/deblending parameters), the software was unable to detect and isolate all identified clumps, leaving some gaps and blending some crowded objects into a single detection. Therefore, we manually select apertures on the basis of the visual inspection of the previous step. We centre each aperture at the brightest pixel of a given clump, using a radius that matches its observed (magnified) size in the F814W band. An example of our selection can be seen in Fig.\,\ref{fig:sys1}.

To search for astrometric anomalies, we compare our identified centroid positions with the \citet{lagattuta2023} lens model predicted positions. In doing so, we find that no clump is more than 0.45\arcsec\ from its expected location, which falls within the global model uncertainty, i.e., 0.50\arcsec\ as given in \citet{lagattuta2023}. This is to be expected, as many of these knots are used as constraints in the model (particularly System\,1) leading to minimized predictions during model optimization. Nevertheless, we note that the largest model residuals are located near Images\,1.2, 1.3 and 2.2, providing a starting point to look for perturbers.

\begin{figure*}
    \centering
    \includegraphics[width=0.24\textwidth]{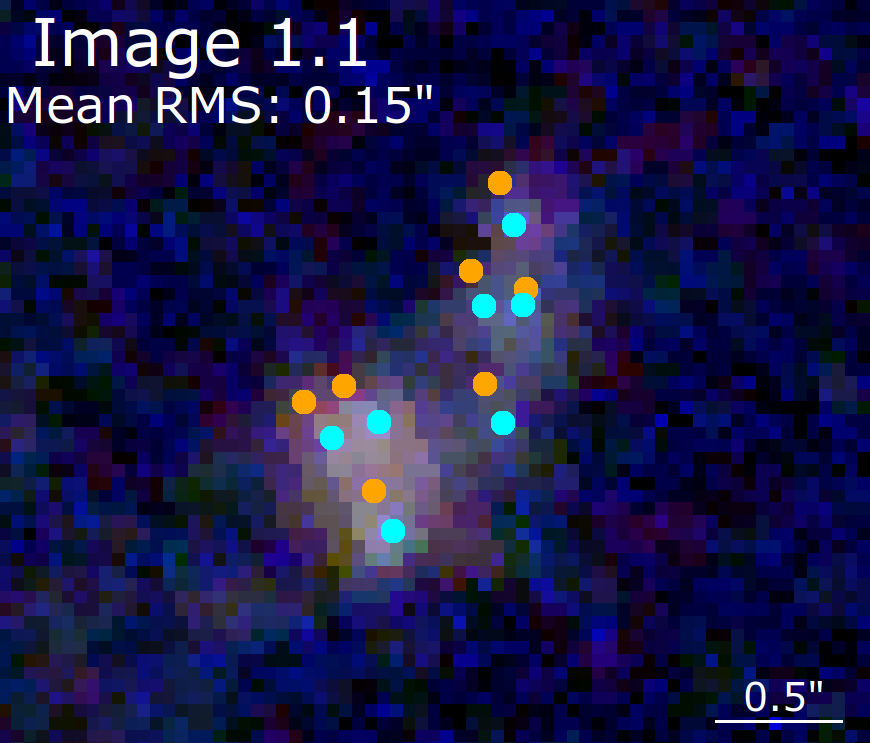}
    \includegraphics[width=0.24\textwidth]{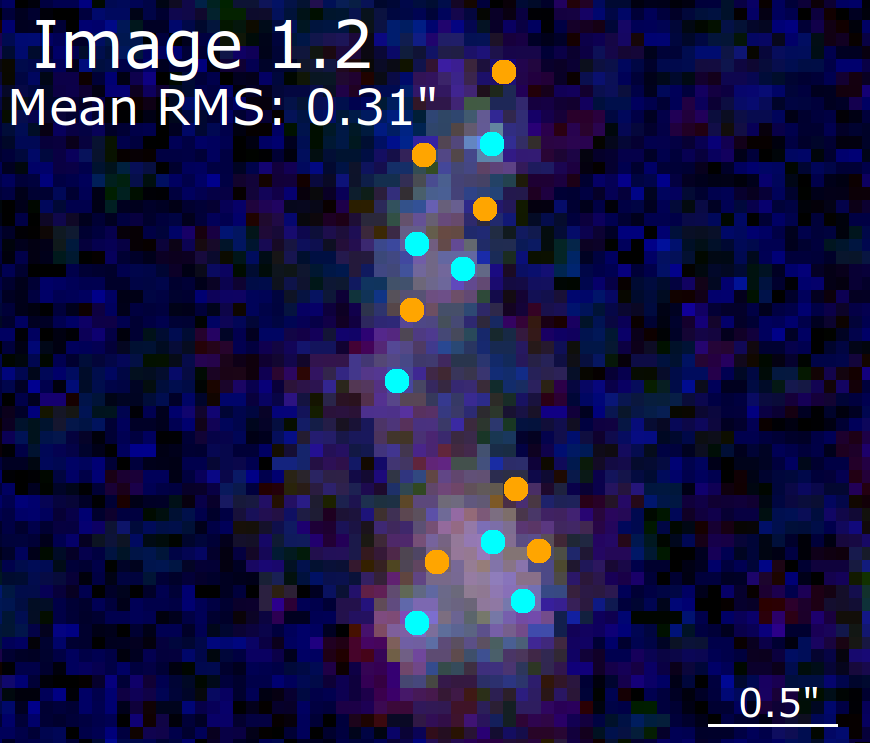}
    \includegraphics[width=0.24\textwidth]{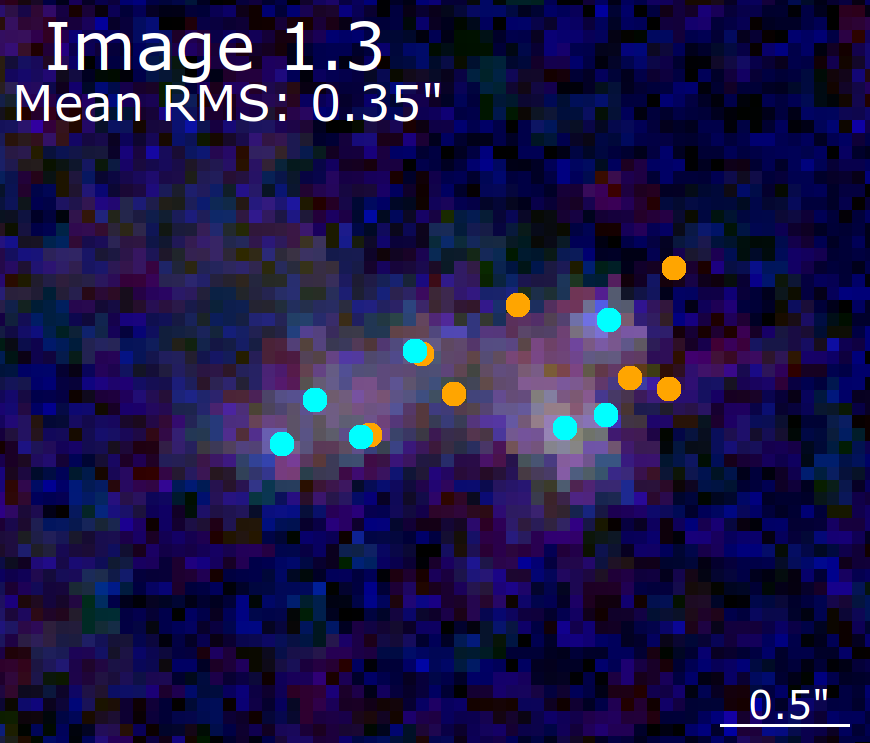}
    \includegraphics[width=0.24\textwidth]{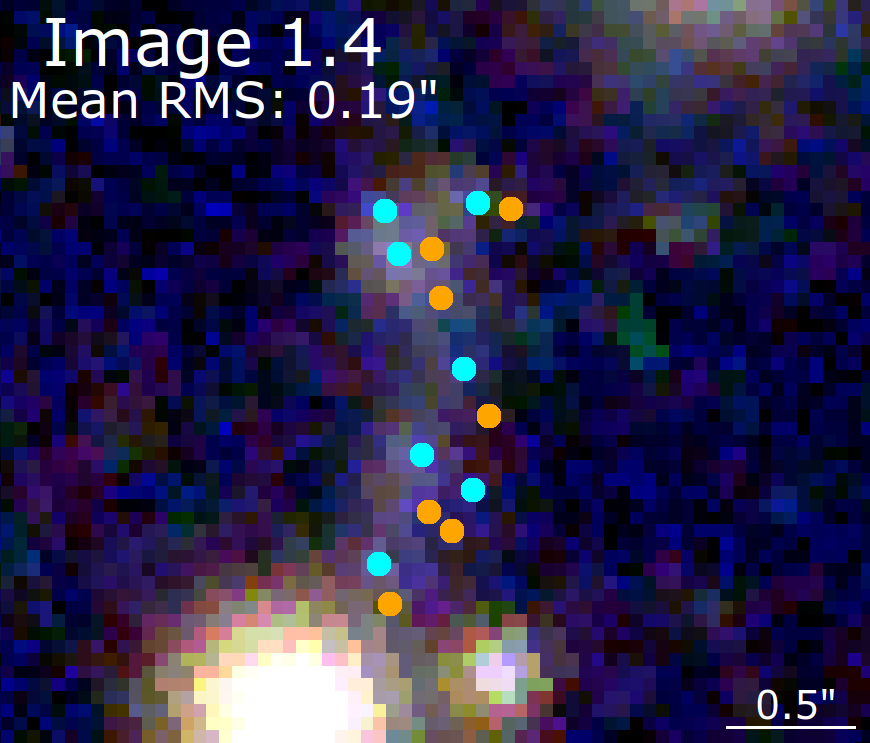}
    \includegraphics[width=0.24\textwidth]{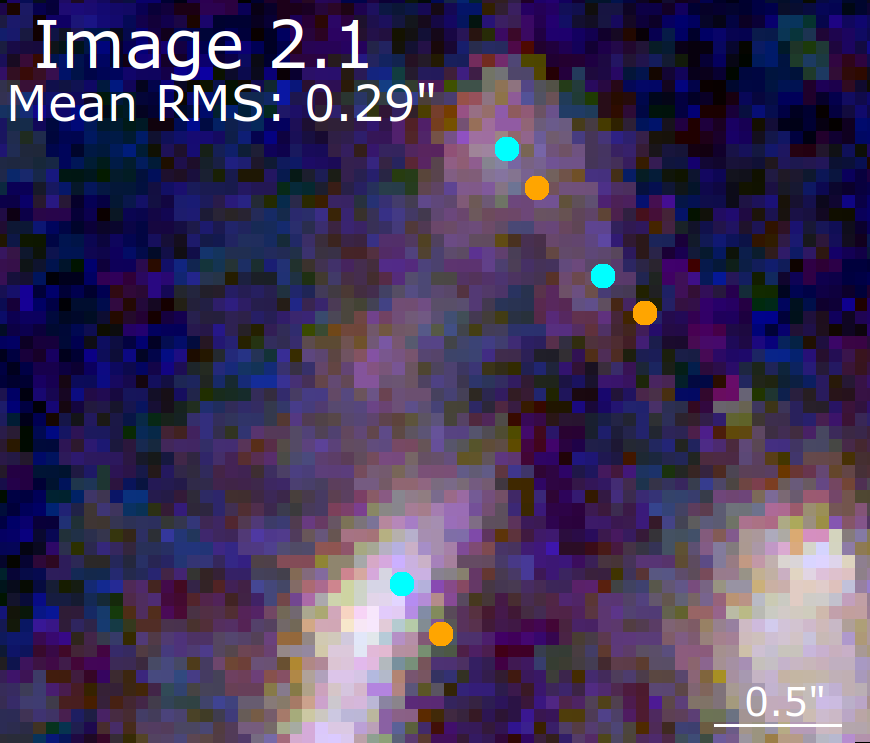}
    \includegraphics[width=0.24\textwidth]{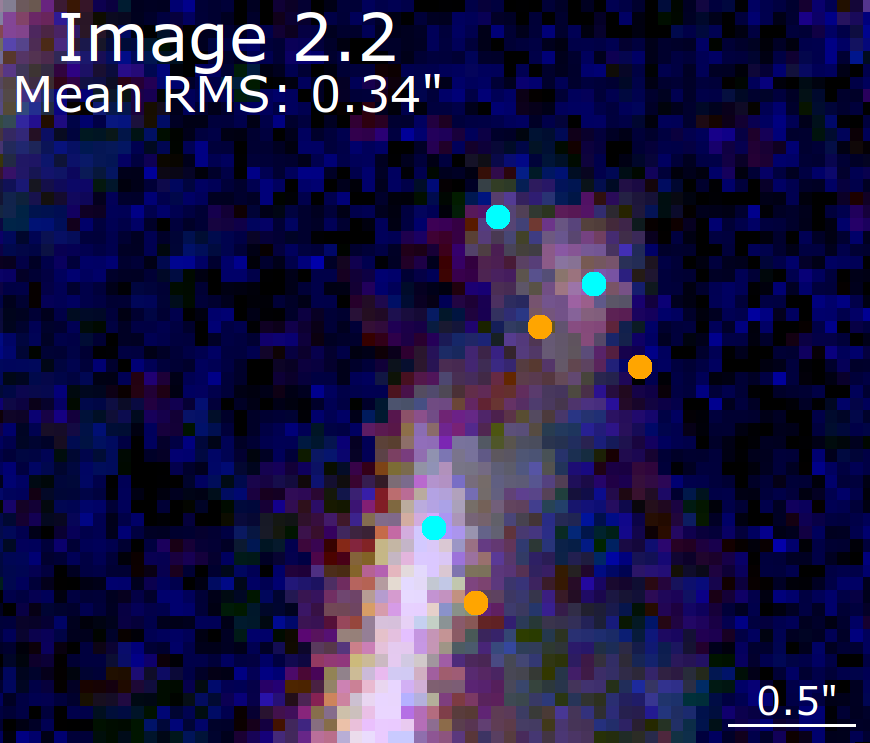}
    \includegraphics[width=0.24\textwidth]{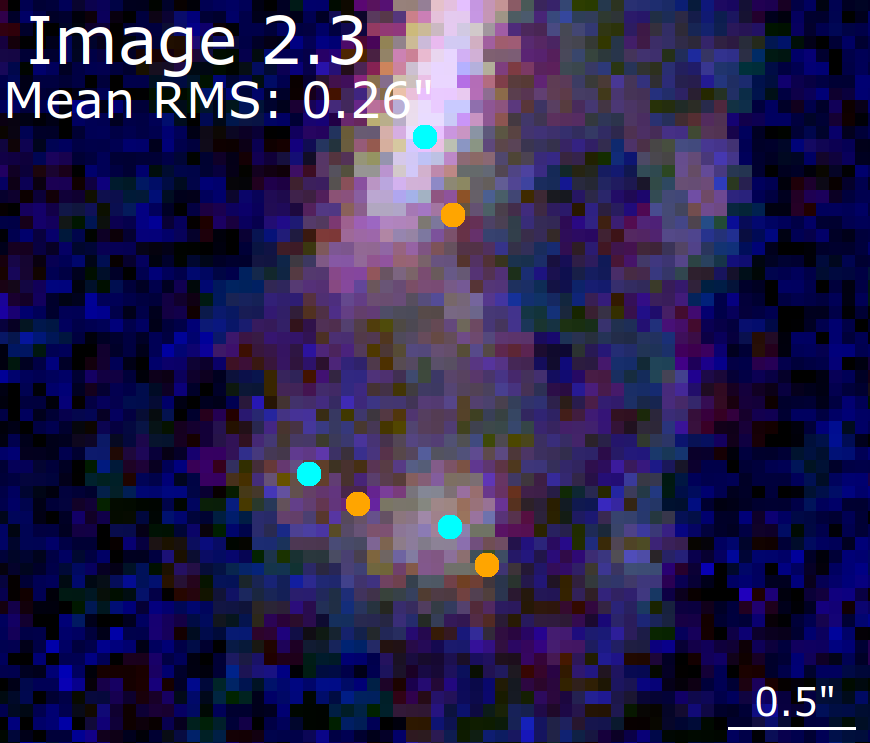}
    \includegraphics[width=0.24\textwidth]{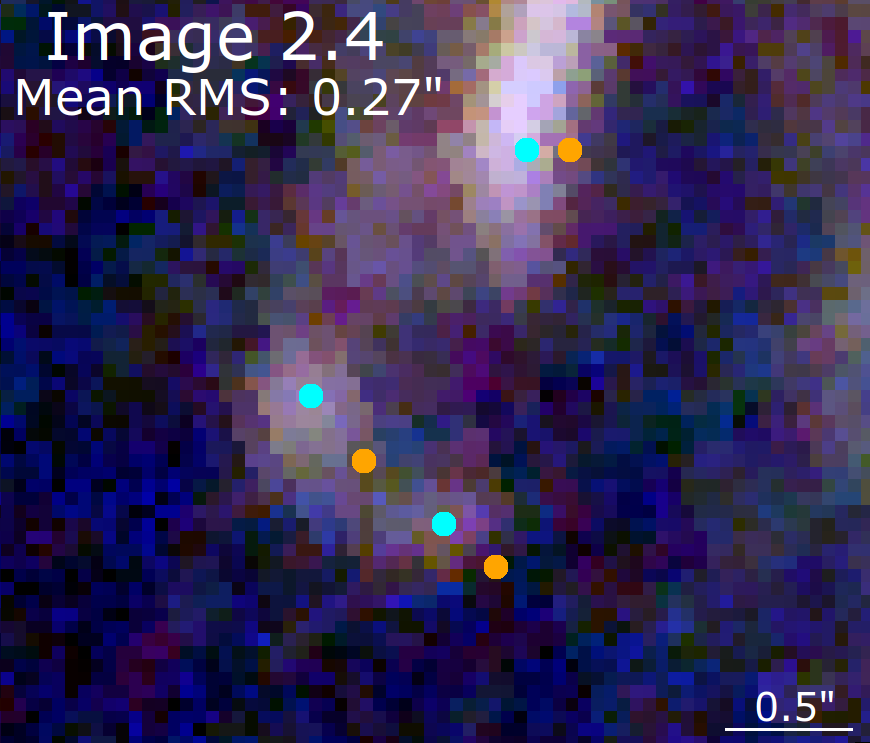}
    \caption{A comparison of the observed stellar clumps in the HU systems (cyan) to the model-predicted positions (orange). The largest discrepancies ($\sim$0.35\arcsec) can be found around Images\,1.2, 1.3, and 2.2.}
    \label{fig:lenstool vs true}
\end{figure*} 

In addition to the astrometric search, we also compare the conjugate fluxes of each clump. Unlike astrometric positions, the clump magnitude is not used as a constraint in the lens model, thus serving as a more independent probe of anomalies. To measure the flux of each clump, we create a self-consistent estimator by summing a scaled version of the flux contained within each aperture. Specifically, we scale the empirically-measured flux in each pixel by its model-defined magnification value. This provides a measurement of intrinsic (i.e. non-magnified) flux, which should remain constant from image to image in the absence of low-mass perturbers. To improve the contrast between clump images, and to minimize any contamination due to cluster light, we first subtract the BCG flux from each filter before measuring aperture fluxes. We model the BCG light using \textsc{MgeFit}, a publicly available python package based on the multi-gaussian expansion (MGE) galaxy fitting technique presented in \citet{capellari2002}. A list of our fitting parameters can be seen in Table~\ref{tab:mgeFit}

\begin{figure*}
    \centering
    \includegraphics[width=0.49\textwidth]{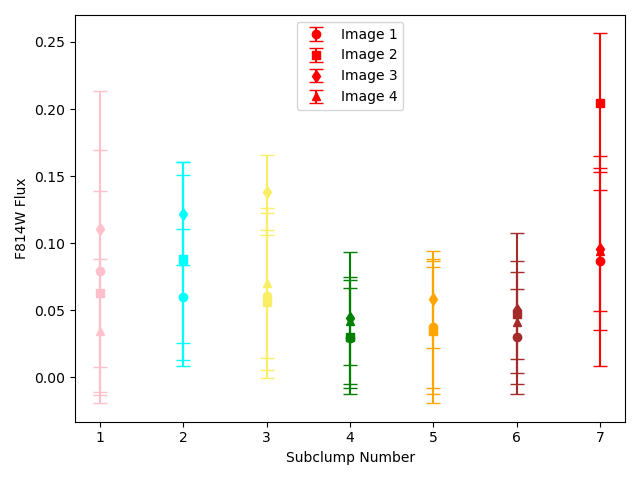}
    \includegraphics[width=0.49\textwidth]{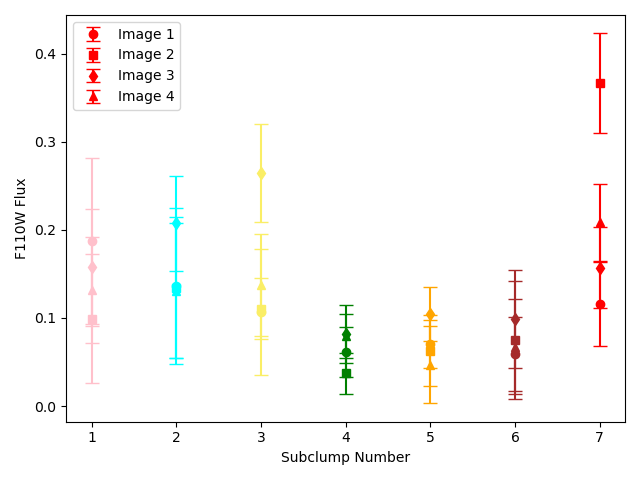}
    \includegraphics[width=0.49\textwidth]{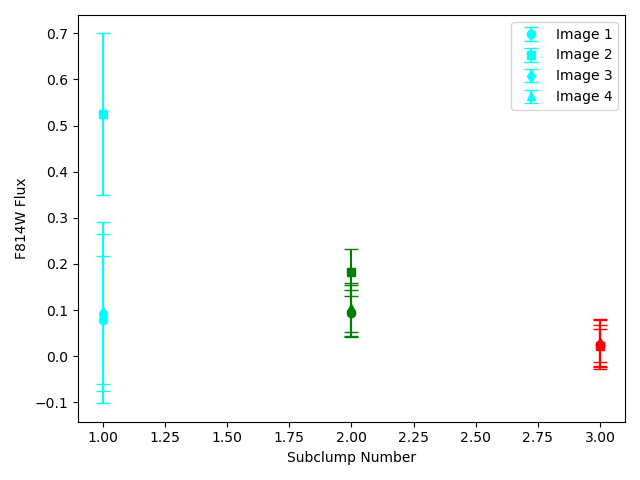}
    \includegraphics[width=0.49\textwidth]{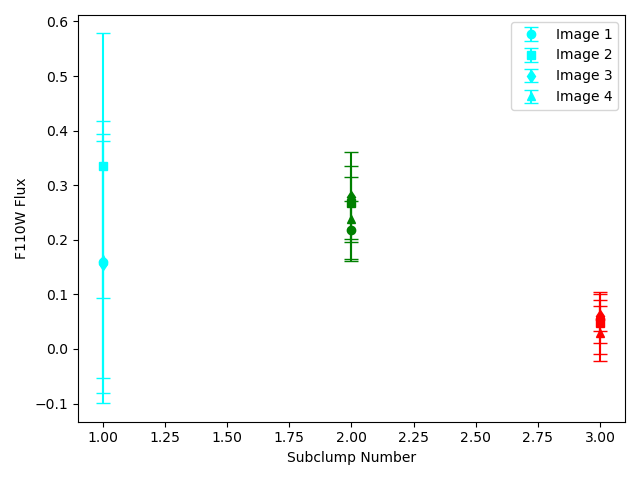}
    \caption{Top panels: Measured aperture fluxes in the F814W and F110W bands for each of the seven distinct sub-clumps identified in System\,1, after correcting for magnification effects. An example of apparently discrepant points can be seen in the F110W plot. Subclump colours correspond to those seen in Fig.\,\ref{fig:sys1}. Bottom panels: same, but for System\,2. Here, subclump colours correspond to those seen in Fig.\,\ref{fig:sys2}.}
    \label{fig:colourAnomalies}
\end{figure*}

We provide estimates for each clump using all three \emph{HST} bands, as well as the MUSE flux of the brightest emission line for each system (in this case, Ly-$\alpha$ for System\,1 and CIII] for System\,2.) The results of these measurements are shown in Fig.~\ref{fig:colourAnomalies}.

\begin{table*}
    \centering
        \caption{Model parameters used during the Multi Gauss Expansion (MGE) suptraction of the BCG. Here, PA is defined in degrees East of North, and ellipticity is given by $\varepsilon = (a^2 - b^2)/(a^2 + b^2)$, where a and b are the semi-major and semi-minor axes of the ellipse, respectively.}
    \begin{tabular}{c|c|c|c|c|c|c}
    \hline
    \hline
        gauss components & psf size & fit mask radius & fit mask PA & fit mask ellipticity & fit mask RA & fit mask Dec\\
         & [\arcsec] & [\arcsec] & [degree] & & [degree] & [degree]\\
         \hline
         15 & 1.8 & 21.75 & 279 & 0.29 & 69.2895902 & 0.7313656\\
     \hline
     \hline
    \end{tabular}
    \label{tab:mgeFit}
\end{table*}

\begin{figure*}
    \centering
    \includegraphics[width=\textwidth]{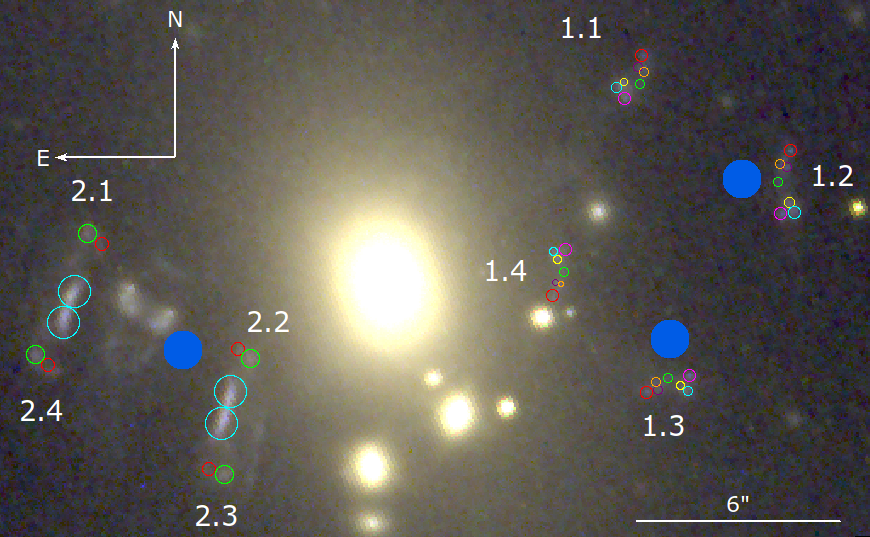}
    \caption{The potential locations of dark matter substructures near Systems\,1 and 2, based on astrometric and flux ration anomalies, are shown by the blue circles. For reference, individual subclumps of each HU image appear as coloured open circles. Once again, rgb colouring is made from F140W/F110W/F814W broadband imaging.}
    \label{fig:possible dark matter locations}
\end{figure*} 

Again, the largest discrepancies are found around Images\,1.2, 1.3, and 2.2, in agreement with the astrometric discrepancies. We therefore consider these locations as our best candidates for searching for possible dark substructures.  

\section{Gravitational lensing mass modelling}
\label{sec:modeling}
To investigate the possible presence of dark substructures in the \rxj\ field, we modify the parameters of the strong lensing mass model presented in \citet{lagattuta2023}. Specifically, we add small perturbing mass components to the original \citet{lagattuta2023} model, and then optimize the added parameters to see if they significantly improve the overall fit. Although there are several lens modelling codes that can do this, we choose to use the parametric modelling software \texttt{Lenstool} \citep{jullo2007}. 

\subsection{Parametrization of substructures} 
\label{subsec:substructures}

We model each DM perturber candidate as a single, compact subhalo, in-line with the appearance of substructures in numerical simulations \citep[e.g.][]{despali2017,benitez-llambay2020}. Following the convention of \citet{lagattuta2023}, we parametrize these subhalos using a Pseudo-Isothermal Elliptical Mass Distribution \citep[PIEMD;][]{eliasdottir2007}. This choice is not arbitrary, as due to the self-similar nature of DM, it is standard to use a common profile for both large and small structures in lensing studies \citep[e.g.][]{jauzac2015, lagattuta2019, caminha2023, furtak2024, allingham2025}. Additionally, a PIEMD halo is physically similar to the more universal NFW profile \citep{nfw1997}, but is computationally faster to evaluate. At the same time, a PIEMD model provides a central flattening of the halo (a ``core'' component), allowing for more control over the mass distribution. 

The PIEMD halo has seven tuneable parameters: centroid position ($\alpha$, $\delta$), ellipticity ($\varepsilon$), position angle ($\theta$), two characteristic radii ($r_{\rm core}$ and $r_{\rm cut}$) that control the inner and outer-most mass distributions (where the halo deviates from a purely isothermal profile), and finally a velocity dispersion ($\sigma$) which, when combined with the characteristic radii, encodes the total halo mass. For larger mass components, i.e., cluster-scale and galaxy-scale halos, several of these components can be set by empirical information. For example, the overall distribution of cluster member galaxies can guide the ellipticity of a cluster halo, while the positions and morphologies of individual cluster members can be observed directly in broadband imaging. 

For our candidate subhalo perturbers, since we do not have any empirical knowledge of their appearance, we make some simplifying assumptions during the mass modelling. First, we assume each halo has spherical symmetry, removing the need to fit $\varepsilon$ and $\theta$. This follows the convention of previous subhalo detection studies \citep[e.g.,][]{vegetti2012, he2022, nightingale2024}, who find that differences between ellipitcal and spherical perturbers on the lensing signal are, as of yet, still too weak to be reliably discerned. For similar reasons, we fix the core radius component of each clump to a small (but non-vanishing) 0.15\,kpc. This increases the compactness of the perturbers, increasing their lensing efficiency while minimizing their mass. In this way, we can treat any obtained values as a lower limit on the derived subhalo mass estimate. This, when combined with an upper limit derived from a (non) detection of a luminous component, constrains the possible phase space of allowable DM mass. After making these assumptions, we are left with only four variable parameters ($\alpha$, $\delta$, $\sigma$, and $r_{\rm cut}$) for each halo, which we optimize independently for each component. 

\subsection{Construction and optimization of the mass model}
\label{subsec:optimization}

\subsubsection{Construction}
After deciding on the form of our candidate perturbers, we include them to the model. Following the abundance of anomalies in the observational data and the original lens model (Section\,\ref{subsec:clumpProperties}), we treat the area around Images\,1.2, 1.3, and 2.2 as the most obvious regions to make our initial guesses. To be conservative (in the absence of any other information), we place a single candidate DM halo in the vicinity of each image. Given the low mass, compact nature of our assumed subhalo components, any effects they have on the global lens model will remain confined to a small area around their positions. Since the three ``most anomalous'' lens images are at a considerable distance from one another, this suggests that individual subhalos should not interact with or affect one another in any significant way, leaving them as independent probes. (Fig.~\ref{fig:possible dark matter locations})

To test this idea, we construct four separate ``substructure-inclusive'' models of the \rxj\ field: three in which we place a single subhalo candidate close to one of the affected multiple images (models 1, 2, and 3) and a fourth (model 4) that includes one subhalo around each image simultaneously. Model 1 focuses on the area around HU Image\,1.2, Model 2 around Image\,1.3, and Model 3 around Image\,2.2. We label the three elements of model 4 as Model 4.1, 4.2, and 4.3, (with, e.g., Model 4.1 corresponding to the subhalo in Model 1). 
We stress, however, that these three components are fit simultaneously and are not considered as separate models. By constructing the models in this way, we can compare the final halo parameters between models more easily, giving us a measure of parameter stability but also an indication of component confidence: if a candidate substructure  remains in a similar position and maintains a similar mass after both its own individual model optimization and the optimization of combined candidates, we can be more confident in its possible existence.

\subsubsection{Optimization}
In order to improve the computational efficiency of our models, rather than re-optimizing all elements, we instead only allow the subhalo components to vary, fixing the remaining (larger-scale) parameters to the final values presented in \citet{lagattuta2023}. This significantly increases the speed of a model run and also prevents the small substructure components from being overwhelmed by larger neighbours. Using \textsc{Lenstool}, we optimize each models (Models 1-4) separately, then compare their parameters after they have converged.

During optimization, we apply moderate constraint limits to each subhalo parameter. This allows us to effectively explore parameter space, while preventing the subhalo components from drifting into non-physical regimes. Specifically, we allow the centroid of each subhalo ($\alpha$,$\delta$) to vary within a 6\arcsec$\times$6\arcsec\ box, centred around the mean coordinates of all observed subcomponents of a given HU image. We similarly constrain the subhalo velocity dispersion, $\sigma$, to vary between 0 and 50\,km/s, and the cutoff radius, $r_{\rm cut}$, between 1 and 10\,kpc \citep[e.g.,][]{mahler2023, doppel2025}. Combined, these constraints maintain the shape of the subhalo as small and compact, and keep it within the effective area of the target HU image while still allowing for the possibility of a non-detection. 

\begin{figure*}
    \centering
    \includegraphics[width=\textwidth]{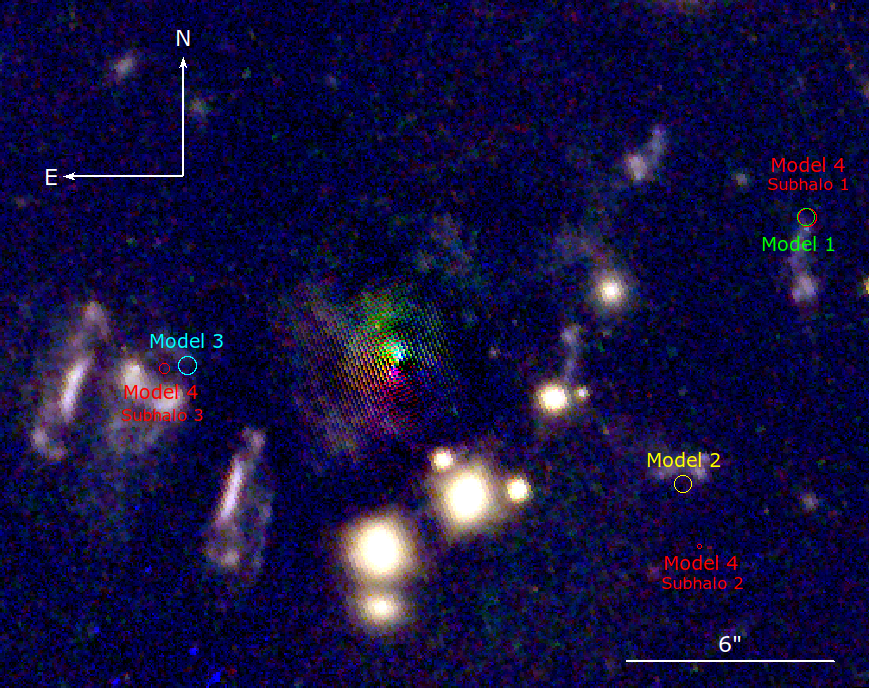}
    \caption{Graphical representation of the best-fit model results where here the BCG has been subtracted to improve the contrast (following the procedure described in Sect.\,\ref{subsec:clumpProperties}. Single-subhalo models are shown in green (model 1), yellow (model 2), and cyan (model 3) respectively, while the multi-subhalo model (model 4) appears in red. The final position of each halo is represented by a circle, with sizes representing a scaled combination of velocity dispersion, which serves as a proxy for halo mass. The 6\arcsec\ scale bar also shows the size of the bounding box over which each subhalo is able to move during optimization. We see that the final parameters for the subhalo around HU Image\,1.2 (model 1 and Model 4, subhalo 1) are entirely consistent, while the other candidate subhalos change significantly between models.}
    \label{fig:best_fit_positions}
\end{figure*}

The optimization process itself is handled by \textsc{Lenstool}, using its built-in fitting routine in the lens plane. After each model converges to its best-fit parameter set, the code robustly explores uncertainties in parameter space, generating 1000 MCMC realizations of each model. The results of this exploration can be seen in Table\,\ref{tab:best fit positions}, and a graphical representation of the final halo candidates can be seen in Fig.\,\ref{fig:best_fit_positions}.

\begin{table}
\centering
\caption{Best-fit values of the candidate subhalos for all four models. Column 1 lists the model considered. Columns 2 and 3 give the best-fit values of Right Ascension and Declination (and their respective errors) in arcseconds, relative to the BCG ($\alpha = 69.2896877$, $\delta = 0.7311409$) at the centre of \rxj. Column 4 presents the dPIE cutoff radius, while column 5 lists velocity dispersion. Model 4 is run with all three subhalo candidates inserted, however, for clarity we list the best-fit position of each individual subhalo. 
}
    \label{tab:best fit positions}
        \begin{tabular}{c|c|c|c|c}
            \hline
            \hline
            Model & $\Delta_\alpha$ & $\Delta_\delta$ & $r_{\rm cut}$ & $\sigma$\\
             & [\arcsec] & [\arcsec] & [kpc] & [km/s]\\
            \hline
            Model 1 & $11.87^{+3.47}_{-0.47}$ & $4.20^{+1.79}_{-1.86}$ & $1.45^{+0.16}_{-0.68}$ & 
            $38^{+1}_{-3}$\\
           %  \hline
            \rule{0pt}{3ex} Model 2 & $8.33^{+2.70}_{-0.65}$ & $-3.51^{+0.54}_{-1.80}$ &
            $2.91^{+0.28}_{-0.95}$ & 
            $38^{+1}_{-7}$\\
             %\hline
            \rule{0pt}{3ex} Model 3& $-5.53^{+1.25}_{-0.75}$ & $-0.10^{+1.30}_{-1.19}$ &
            $2.97^{+0.21}_{-1.29}$ & 
            $40^{+1}_{-6}$\\
             %\hline
            \rule{0pt}{3ex} Model 4 (Subhalo 1) & $11.91^{+1.42}_{-2.50}$ & $4.19^{+2.41}_{-1.93}$ &
            $0.68^{+0.09}_{-0.25}$ & 
            $40^{+1}_{-2}$\\
            %\hline
            \rule{0pt}{3ex} Model 4 (Subhalo 2) & $8.79^{+2.26}_{-0.13}$ & $-5.34^{+0.89}_{-1.07}$ &
            $2.52^{+0.29}_{-1.09}$ & 
            $10^{+2}_{-6}$\\
            %\hline
            \rule{0pt}{3ex} Model 4 (Subhalo 3) & $-7.67^{+1.31}_{-1.57}$ & $-0.17^{+1.32}_{-0.69}$ &
            $1.02^{+0.13}_{-0.72}$ & 
            $23^{+5}_{-1}$\\
            \hline
            \hline
        \end{tabular}
\end{table}

\subsection{Model results} 
\label{subsec:results}

Looking at the final outcomes, we can compare the best-fit parameters of the individual halo models (models 1, 2 and 3) to the combined model (model 4). Doing this, we see that the subhalo positions around Images\,1.3 and 2.2 move significantly between cases, with centroids that are inconsistent within uncertainty measurements. At the same time, the velocity dispersion parameters between models also diverge, and, in some cases, the values are consistent with zero. This suggests that either models have not converged, or there is not enough information to truly constrain a subhalo here. Given the wide-ranging MCMC exploration in the previous step, non-convergence seems unlikely, making it more probable that the signal is simply not strong enough to convincingly reveal the presence of a subhalo.  

Conversely, the two models that include HU Image\,1.2 show good agreement between them, with overlapping centroid positions and a velocity dispersion that is internally consistent and also inconsistent with zero. This provides stronger evidence for the existence of a subhalo which warrants further exploration. Specifically, we compare the fit statistics of model 1 to the original one presented by \citet{lagattuta2023}, finding that the new model has both a smaller rms and lower total chi$^2$. Combining these pieces of information together, Image\,1.2 subhalo becomes our best DM candidate.

Finally, we can convert the model parameters to a physical mass value. For a PIEMD profile, this procedure is described in Appendix A7 of \citet{eliasdottir2007}:

\begin{equation} \label{eq:original mass}
    \begin{split}
        \sigma^2(r_{\rm core}) & = \frac{2G}{\Sigma(r_{\rm core})} \times \\
        & \int_{r_{\rm core}}^{\infty} \frac{M(r_{\rm cut})\rho(r_{\rm cut})}{r_{\rm core}^2}\sqrt{r_{\rm cut}^2 - r_{\rm core}^2}dr_{\rm cut}
    \end{split}
\end{equation}

which, when solved and rearranged becomes:

\begin{equation} \label{eq:rearranged mass}
    M(r_{\rm cut}) = \frac{\sigma^2(r_{\rm core})}{G*\frac{r_{\rm cut}}{r_{\rm cut}^2 - r_{\rm core}^2}*\frac{1}{2\pi}*\frac{4}{3}}
\end{equation}

Substituting in the values from model 1, we measure a final subhalo candidate mass of $(2.25 \pm 0.94) \times 10^9 \: M_\odot$. A full list of masses for all subhalos over all models can be found in Table\,\ref{tab:model masses}.

\begin{table}
    \centering
        \caption{Best-fit masses for all substructure models.}
    \begin{tabular}{|| c | c ||}
    \hline
    \bf Model & \bf Mass ($M_\odot$) \\
    \hline\hline
    Model 1 & $(2.25\pm 0.94) \times 10^9$ \\ 
    \rule{0pt}{3ex}
    Model 4 (Subhalo 1) & $(2.15\pm0.64) \times 10^9$ \\
    \rule{0pt}{3ex}
    Model 2 & $(4.69\pm0.17) \times 10^9$ \\
   \rule{0pt}{3ex}
    Model 4 (Subhalo 2) & $(2.92\pm0.15) \times 10^8$ \\
   \rule{0pt}{3ex}
    Model 3 & $(5.19\pm0.98) \times 10^9$ \\
   \rule{0pt}{3ex}
    Model 4 (Subhalo 3) & $(5.79\pm0.36) \times 10^8$ \\
    \hline
    \end{tabular}
    \label{tab:model masses}
\end{table}

\section{Discussion}
\label{sec:discussion}

\subsection{Reliability of detections}
\label{subsec:reliability}

The possible detection of an invisible mass perturber in a galaxy cluster is an intriguing possibility, but we stress that this result is preliminary and care must be taken to interpret it. In particular, we note that the \emph{HST} imaging data we have used for this analysis is shallow, limiting our ability to detect and discriminate between perturbers. At the same time, the total mass of the candidate subhalo ($\sim 2 \times 10^9 \: M_\odot$), while many orders of magnitude smaller than the cluster halo, is on the higher side of potential subhalo detections \citep[e.g.,][]{vegetti2012, hezaveh2016, minor2021, powell2025}, lessening its ability to discriminate between DM models, as significant deviations only become apparent at cluster masses $< 10^8 M_{\odot}$. At the same time, a more detailed study and analysis will require a statistical approach, based on several halo detections. Nonetheless, this work represents an important first-step in detecting low mass deflectors in the densest regions of galaxy clusters, increasing the likelihood of finding other examples, with more varied halo masses, in the future. 

\subsection{Is the subhalo candidate truly ``dark''?}
\label{los_interlopers}
Along similar lines, it is important to note that, inherently, our analysis has assumed that the subhalo perturbers are embedded within the main cluster halo. However, this is not the only plausible scenario: it is also possible that these objects lie along the line of sight (LoS). Cosmological simulations have previously shown that, while a cluster dominates the mass budget of its line of sight, the distribution of objects responsible for the observed lensing signal has a much broader distribution \citep{robertson2020}. 

This can be especially problematic for detecting and distinguishing DM perturbers, as these cannot be reliably redshifted, regardless of imaging quality. Indeed, if this is the case here, it can significantly change our interpretation of the result: a more distant LoS deflector would have a different mass and a lower observable flux density. Combined, this could make the deflector a less viable candidate for discriminating between DM models. Specifically, a more massive, distant object would be even less sensitive to differences between CDM and its alternatives, and may not be truly dark -- rather, luminous but below the detectable flux limit.

Importantly, recent work has shown that the observable effects of LoS deflectors are sensitive to image depth, with deeper imaging more sensitive to these interlopers than shallow imaging \citep{he2022}. Given the shallow nature of our current imaging data, we are likely less sensitive to LoS perturbers, and can, at the moment, attribute any distortions to a subhalo in the cluster field itself. Nevertheless, the possibility of interlopers remain. To truly confirm or reject this possibility, we must turn to deeper and higher-resolution data.

\section{Conclusions and future work}
\label{sec:conclusions}

In this work, we have studied mass distributions in the galaxy cluster \rxj, taking advantage of three high-magnification, ``exotic'' Hyperbolic-Umbilic gravitational lenses (designated Systems\,1, 2, and 10 in \citealt{lagattuta2023}) to search for low-mass ($m \lesssim 10^9 M_{\odot}$) subhalos in the cluster core. We summarize our results as follows:

\begin{itemize}
    \item Using a combination of high-resolution \emph{HST} imaging and VLT/MUSE spectroscopy, we identify seven distinct luminous subcomponents (clumps) in System\,1, three in System\,2, and a single bright Lyman-$\alpha$ emission region in System\,10.

    \item After subtracting contamination due to the nearby BCG and correcting for lensing magnification, we compare the brightnesses between the multiply-imaged clumps of each system (which should ostensibly be identical). Doing so, we find that segments of Images\,1.2, 1.3, and 2.2 are noticeably different from their counterparts.

    \item We also compare the observed positions of each HU clump to its predicted location in the \citet{lagattuta2023} lens model, again finding the largest discrepancies in Images\,1.2, 1.3, and 2.2. Together with the luminosity discrepancies, we focus on these regions for our substructure search.

    \item To investigate the presence of substructure, we embed several candidate dark matter subhalos in the \citet{lagattuta2023} lens model, using the parametric modelling software \textsc{Lenstool}. We specifically add these candidates in the vicinity of Images\,1.2, 1.3, and 2.1.

    \item After optimizing the new lens models, we see that one candidate subhalo (the object in the region of Image\,1.2) maintains a consistent position and total mass: $(2.15 \pm 0.94) \times 10^9 M_{\odot}$ which we consider as our best possible substructure candidate.

\end{itemize}

While potentially interesting, we stress that this detection is still preliminary: our best available imaging data are shallow (\emph{HST} SNAPshot images) resulting in lower confidence measurements. To gain confidence in the result, we will need deeper and higher resolution data. This will soon be available in the form of \emph{James Webb Space Telescope} (\emph{JWST}) data (PID: 6207, PI D.~Lagattuta), allowing us to explore this cluster further. However, with this work, we demonstrate the feasibility of using galaxy cluster lenses, and their HU systems, as laboratories to study the distribution of low mass substructures.

Besides acquiring further data however, we are also endeavouring to expand the sample of known HU lenses. This will allow us to investigate DM in a statistical way, using a sample of subhalo detections to place more robust constraints on the low-mass end of the subhalo mass function. To that end, we have already identified additional, promising HU candidates, both in archival \emph{HST} data and in recent \emph{JWST} imaging -- particularly the Strong LensIng and Cluster Evolution (SLICE) sample: a collection of 124 galaxy clusters observed with \emph{JWST} as part of a survey programme (PID: 5594; PI G.~Mahler; \citealt{cerny2025}). We will continue to pursue these efforts in follow-up work.

\section*{Acknowledgements}

%The Acknowledgements section is not numbered. Here you can thank helpful
%colleagues, acknowledge funding agencies, telescopes and facilities used etc.

DJL, MJ, and JED acknowledge support by the United Kingdom Research and Innovation (UKRI) Future Leaders Fellowship `Using Cosmic Beasts to uncover the Nature of Dark Matter' (grant number MR/X006069/1).

%%%%%%%%%%%%%%%%%%%%%%%%%%%%%%%%%%%%%%%%%%%%%%%%%%
\section*{Data Availability}

The inclusion of a Data Availability Statement is a requirement for articles published in MNRAS. Data Availability Statements provide a standardised format for readers to understand the availability of data underlying the research results described in the article. The statement may refer to original data generated in the course of the study or to third-party data analysed in the article. The statement should describe and provide means of access, where possible, by linking to the data or providing the required accession numbers for the relevant databases or DOIs.



\bibliographystyle{mnras}
\bibliography{substructurePaper} % if your bibtex file is called example.bib

%\begin{thebibliography}{99}
%\end{thebibliography}

%%%%%%%%%%%%%%%%%%%%%%%%%%%%%%%%%%%%%%%%%%%%%%%%%%

%%%%%%%%%%%%%%%%%%%%%%%%%%%%%%%%%%%%%%%%%%%%%%%%%%

% Don't change these lines
\bsp	% typesetting comment
\label{lastpage}
\end{document}